\documentclass[12pt,preprint]{aastex}

\shortauthors{Carciofi, Bjorkman and Magalh\~aes}

\begin{document}

\title{Effects of Grain Size on the Spectral Energy Distribution of Dusty Circumstellar Envelopes}

\author{A. C. Carciofi}
\email{acarcio@physics.utoledo.edu}
\author{J. E. Bjorkman}
\affil{Ritter Observatory, Dept. of Physics and Astronomy,
University of Toledo, Toledo, OH 43606-3390} \and
\author{A. M. Magalh\~aes}
\affil{Instituto de Astronomia, Geof\'{\i}sica e Ci\^encias
Atmosf\'ericas, Universidade de S\~ao Paulo, Caixa Postal 3386
01060-970, S\~ao Paulo, Brazil}

\begin{abstract}

We study the effects of dust grain size on the spectral energy distribution (SED) of spherical circumstellar envelopes. Based on the self-similarity relations of dusty SEDs derived by \citet{ive97}, 
we expect an approximate invariance of the IR SED for models with different grain sizes.
Approximate invariance follows from the fact that differently sized grains have similar optical properties at long wavelengths where the dust reprocesses the starlight. In this paper, we discuss what are the physical requirements on the model parameters to maintain the approximate invariance of the IR SED.  Single grain size models are studied for a wide range of grain sizes in three optical depth regimes: optically thin models, moderate opacities, and very optically thick models. In this study, we find limits for the cases where the IR SED is and is not capable of conveying information about grain sizes, and to what extent it does so. 
We find that approximate invariance occurs for a much larger range of grain sizes than previously believed, and, when approximate invariance holds, the SED is controlled mainly by one parameter, the reprocessing optical depth, a quantity that measures the fraction of starlight that is absorbed by the dust grains.
Models with a grain size distribution are studied as well.  For these models, we find that, in many instances, the concept of approximate invariance may be extended from the IR SED to all wavelengths.
This means that, for a wide range of optical depths, models with different grain size distributions will produce very similar SEDs and, hence, the reprocessing optical depth is the only quantity that can be unambiguously obtained from the SED.
The observational consequences of this result are discussed in detail. Finally, in models with a size distribution, the different grain sizes each have different equilibrium temperatures. The consequences of this effect for the model SED are discussed as well.
\end{abstract}


\section{Introduction \label{introduction}}

Dust is associated with many astronomical objects, such as stars and galaxies. Dust grains typically absorb radiation at short wavelengths, and since the dust grains are usually cooler than the radiation sources, they reemit the absorbed radiation at longer wavelengths. Consequently, the presence of dust is usually  detected by a flux excess at infrared wavelengths. The usual challenge is to use the spectral energy distribution (SED) of the object, to extract information about the nature of the underlying luminous source and the properties of the surrounding dust; i.e., its spatial distribution, optical depth and intrinsic properties. 
The intrinsic properties of the dust grains are their chemical composition, condensation temperature, shape and size. 

Knowledge of the dust grain sizes present in circumstellar matter is desirable for many astrophysical situations. For example, it is believed that about 60\% of the dust injected into the ISM originates from the winds of AGB stars \citep{geh89}, so knowledge of the size of the dust grains present in the winds of such stars is fundamental for understanding the properties of interstellar dust. Similarly, grain size is important for understanding the details of grain formation and mass loss mechanisms for cool stars. If we assume that the source spectrum and the spatial distribution of the dust are known for these objects, the question becomes to what extent is it possible to obtain the intrinsic properties of the dust from knowledge of the SED alone. 
 In this paper, we investigate how grain size affects the SED and to what extent the SED can be used to constrain the grain sizes.

It is well known that models of the SED are not unique. For this reason it is imperative to identify the fundamental parameters controlling the SED. Recently, \citet{ive97}, hereafter IE97, investigated the scaling properties of the radiation transfer problem in spherically symmetric dust-envelopes. 
They found that the radiation transfer problem in dusty media depends only on the following quantities:

i) The geometry (shape) of the system (i.e., all distances are proportional to a scale factor);

ii) The sublimation temperature of the material;

iii) The spectral shape of the input radiation, $\lambda F_\lambda/F$;

iv) The shape of the dust absorption and scattering opacities, $\kappa_\lambda/\kappa_{\lambda_0}$ and $\sigma_\lambda/\sigma_{\lambda_0}$, respectively, where $\lambda_0$ is the fiducial wavelenght;

v) The dust scattering phase function (SPF)\footnote{This condition was not required by \citet{ive97} since they only considered isotropic scattering. We consider dust with anisotropic phase function, therefore this additional requirement is necessary.};

vi) The overall optical depth at the fiducial wavelength.

The above list can be regarded as a set of  {\it invariance requirements}. Two models that have different physical parameters, but meet the above requirements exactly, are equivalent and have the same SED. For example, the physical dimensions of the dusty region can be freely changed (say by changing the stellar luminosity, which increases the dust condensation radius) with no effect on the SED, as long as the overall optical depth and shape factors do not change.
If an invariance requirement is violated, however, then the models are no longer equivalent and the SEDs are expected to be different.
For instance, if one were to change the grain size, the shape of the opacity and the SPF would change, violating requirements (iii) and (v).

For a given class of astronomical objects (e.g., AGB stars), most of the quantities above are likely to be similar (geometry and grain composition), so the optical depth becomes the single most important parameter that controls the SED.

It is reasonable to expect that if an invariance requirement is weakly violated, the SED will still be approximately the same. This was noted by IE97 who pointed out that if the grains are very small (about a tenth of wavelength of the peak of the source spectrum) then the shape of the opacities are very similar and the grain size is irrelevant for the problem. For a 3000 K source, for example, the upper limit for the grain size is about $0.05 \mu \rm m$. Grains larger than this upper limit will, according to IE97, significantly alter the results.

In the case of optically thin envelopes, this condition on the maximum grain size can be further relaxed, because the star completely dominates the optical SED while the grains produce the IR SED.  For the IR SED to remain similar between models with different grain radii, it is evident that both the shape of the IR emissivity and the reprocessed luminosity must be the similar. 
For the shape of the IR emissivity to be the similar, it is necessary that the grains have similar absorption efficiencies {\it only in the IR}; this relaxed requirement increases the maximum grain size to at least  $15 \mu \rm m$, assuming the grains have equilibrum temperatures of about 1000 K. 
The condition that the reprocessed luminosity be similar is equivalent to imposing a condition on the optical depth (see next section).

This example shows that, under certain circumstances, it is possible to {\it relax the invariance requirements and still maintain the invariance of the SED}, at least approximately. 
We call this concept  {\it approximate invariance}, and in this paper we demonstrate how it reveals the 
similarities between models with different grain sizes.
In particular, we  study to what extent the grain sizes can be changed, while maintaining the approximate invariance of the SED under a wide range of model parameters. For each case presented, we identify the primary parameter (like optical depth) that controls the SED.

In the next section, we explore in more detail the physical requirements for approximate invariance when the dust grain size is varied.  In section~\ref{mc}, we briefly describe the Monte Carlo code used for the calculations. In section~\ref{single}, we present the effects of grain size on the SED for single grain size models. Section~\ref{dist} extends the study to models with a grain size distribution, and a discussion and summary of the results is presented in section~\ref{discussion}.

\section{Approximate Invariance \label{as}}

Consider an astrophysical system, which consists of a luminous source (e.g., a star) surrounded by a dusty envelope. The physical description of this system must include both the shape and spectrum of the source and the dust properties (chemical composition, grain size, spatial distribution and optical depth). One characteristic of such a model is the presence of an inner cavity.
Typically, for a star that is losing mass (e.g., an AGB star), the location of the inner cavity will be controlled by dust sublimation/condensation. In contrast, a star that has stopped losing mass (e.g., a planetary nebulae) may have a much larger cavity, with a correspondingly cooler radiative equilibrium temperature at the inner edge of the cavity.

Now consider different models of the system, where we vary only three parameters: the dust grain size ($a$), the cavity inner radius ($r_{\rm i}$), and the optical depth ($\tau$). 
Such models violate some of the invariance requirements of section~\ref{introduction}, thus they should have different SEDs. However, it follows from the idea of approximate invariance (AI) that their IR SED will be similar, 
provided that the three conditions bellow are satisfied:

1. The shape of the absorption efficiency factor for the different grain sizes is similar in the spectral range where the grains emit most of their thermal flux;

2. The integrated (bolometric) IR luminosity is the same; and

3. The temperature of the grains at the inner edge of the envelope is the same for all grain sizes.

\begin{figure}[bp]
\plotone{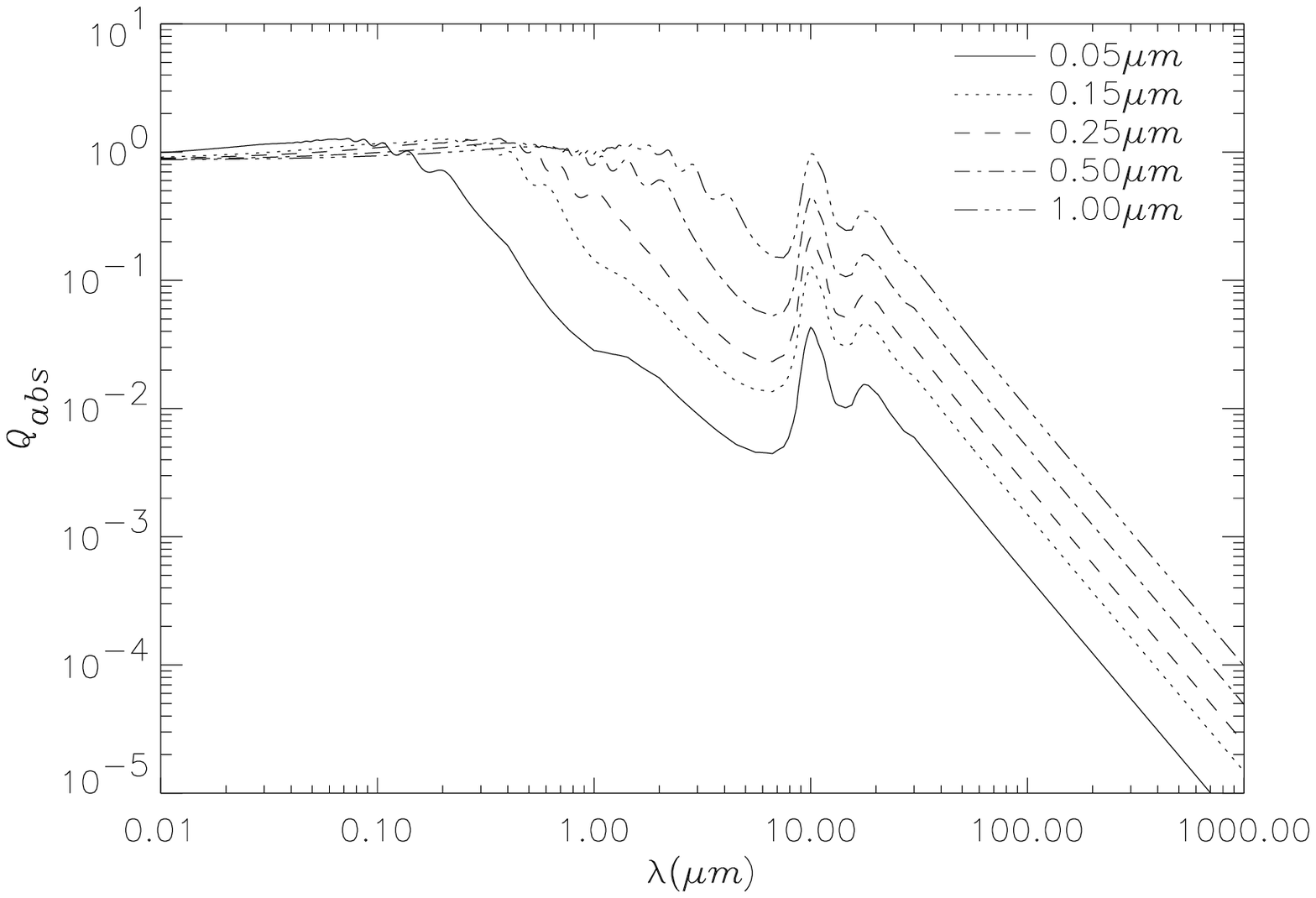}
\figcaption[]{Absorption efficiency factors for spherical cosmic silicate grains. Each line shows $Q_{\rm abs}$ for a given grain radius, as indicated. Optical data for the warm version of \citet{oss92}.
\label{effsil}}
\end{figure}

In the following, we investigate the requirements these conditions impose on the three model parameters (optical depth, grain size and cavity radius). Let us first establish the spectral region where we expect the absorption efficiencies $Q_{\rm abs}$ of differently sized grains to be approximately equal (condition 1). Figure~\ref{effsil} shows $Q_{\rm abs}$ for cosmic silicate of various grain sizes (optical data from \citet{oss92}). Mie theory defines three spectral regions where the grains have common properties:

i) $\lambda \lesssim \pi a$. In this region, called the geometrical limit, the optical properties are roughly independent of $\lambda$, and  $Q_{\rm abs} \approx \rm const.$;

ii) $\lambda \gtrsim 5\pi a$. In this interval, the efficiencies are in the so-called Rayleigh limit, where the size of the particle is much smaller than the wavelength. In this limit, $Q_{\rm abs} \propto 1/\lambda$ for amorphous grains and $Q_{\rm abs} \propto 1/\lambda^2$ for crystalline grains;

iii) $\pi a \lesssim \lambda \lesssim 5\pi a$. In this region, the efficiencies are calculated using Mie theory (spherical particles) and are dependent on grain size. We call this interval the {\it Mie region}.

These three regions are easily recognizable in Figure~\ref{effsil}. We see that all grains have approximately the same efficiencies for short wavelengths (region i) and that the shape of the efficiency factors are the same for $\lambda \gtrsim 10 \,\mu \rm m$ (region ii). In the visible and NIR region, the efficiencies are very different.
From the above discussion, it is easy to see that condition 1 of AI is satisfied when most of the IR radiation is emitted in the Rayleigh limit. In practice, this sets a constraint on the maximum grain size that satisfies condition 1 of AI. For a given temperature of the inner cavity, $T_{\rm i}$, the largest grain size satisfying condition 1 of AI is given by
\begin{equation}
a \lesssim 185/T_{\rm i}\,\mu \rm m. \label{eq1}
\end{equation}
For example, for $T_{\rm i}=1000 \,\rm K$ (approximately the sublimation temperature for most grain species), the maximum grain size satisfying condition 1 of AI is about $0.2 \,\mu \rm m$.

Condition 2 of AI requires that the emergent envelope luminosity, $L_{\rm rep}$, remain fixed. 
The envelope is powered by absorbing some fraction of the stellar luminosity, so condition 2 is equivalent to requiring that this absorbed fraction remain fixed. 
To characterize this condition, we introduce the quantity  $\tau_{\rm rep}$, the {\it reprocessing optical depth}. This quantity is defined so that envelopes with the same $\tau_{\rm rep}$ will reprocess (i.e., absorb) the same fraction of the input energy. The reprocessing optical depth is thus given by the condition
\begin{equation}
1-e^{-\tau_{\rm rep}}\equiv L_{\rm rep}/L_\star \ . \label{trep}
\end{equation}

To define the quantity $L_{\rm rep}$ it is useful conceptually to divide the emergent radiation into stellar photons and envelope photons. 
In a Monte Carlo simulation, the stellar luminosity is divided into $N$ equal energy photon packets (note that packets with different frequencies contain a different number of physical photons). The energy per packet is given by
\begin{equation}
E_\gamma = L_\star \Delta t / N \ ,
\end{equation}
where $L_\star$ is the stellar luminosity and $\Delta t$ is an arbitrary simulation time.
Stellar photons are emitted from the star and they propagate through the envelope, where they may be scattered or absorbed. If a stellar photon is absorbed, radiative equilibrium requires that it be reemitted as an envelope photon. Note that scattering does not change the photon type (stellar vs. envelope). 
The fraction of the stellar luminosity reprocessed (absorbed) by the envelope may easily be determined by counting the number of stellar photon packets absorbed by the envelope, $N_\star^{\rm abs}$. This is equivalent to counting the number of stellar photons that emerge from the envelope without absorption, $N_\star^{\rm em}$. Thus the luminosity reprocessed by the envelope is
\begin{equation}
L_{\rm rep} = N_\star^{\rm abs} E_\gamma/\Delta t  = (N-N_\star^{\rm em}) E_\gamma/\Delta t = L_\star-L_\star^{\rm em}   \ ,
\end{equation}
where $L_\star^{\rm em}$ is the emergent stellar luminosity. This is the method we use to measure $L_{\rm rep}$ in this paper.
Note that scattering affects $L_{\rm rep}$ because a photon packet that is scattered can be absorbed subsequently. Owing to these scattering effects, it is not straightforward to obtain $L_{\rm rep}$ from standard radiative transfer quantities. The reprocessed luminosity is given by the emergent luminosity of the envelope, so
\begin{equation}
L_{\rm{rep}} = \int \left[ 
4\pi j_\lambda + 
\int \frac{d \sigma}{d\Omega}\left(\hat{\rm{n}},\hat{\rm{n}}^\prime \right) 
I_{\lambda}^{\rm{env}} \left( \hat{\rm{n}}^\prime \right) d\Omega^\prime
\right] 
\rho~
e^{-\tau_\lambda(\hat{\rm{n}})}
d\Omega d\lambda dV 
 \ , \label{lenv}
\end{equation}
where  $d \sigma/d\Omega$ is the differential scattering cross section, which is a function of both incoming and outgoing directions ($\hat{\rm{n}}^\prime$ and $\hat{\rm{n}}$, respectively), and $I_{\lambda}^{\rm{env}}$ is the envelope contribution (Monte Carlo envelope photons) to the specific intensity (this includes scattered envelope photons but does not include scattered stellar photons). Energy conservation requires that $L_{\rm rep}$ equals the energy absorbed in the envelope, so $L_{\rm rep}$ can also be written as
\begin{equation}
L_{\rm rep} = 4 \pi \int \kappa_\lambda \rho J^\star_\lambda d\lambda dV \ , \label{lenv2}
\end{equation}
where $J^\star_\lambda$ is the stellar contribution to the mean intensity, including scattered starlight. 

For low optical depths in the point source approximation ($r_i \gg R_\star$), $J_\lambda  \rightarrow  H^\star_\lambda$, where $H^\star_\lambda$ is the source flux. 
Defining $H^\star \equiv \int_{0}^{\infty} H^\star_\lambda d\lambda$ and the {\it flux mean opacity}
\begin{equation}
\kappa_{\rm F} = {\int_{0}^{\infty} \kappa_\lambda H^\star_\lambda d\lambda \over H^\star} \ , \label{sigp}
\end{equation}
eq.~(\ref{lenv2}) can be rewritten, in the optically thin limit, as
\begin{equation}
L_{\rm rep} = 4\pi \int \kappa_\lambda \rho H^\star_\lambda d\lambda ~ 4\pi r^2 dr =
16\pi^2 \int  \kappa_{\rm F} \rho H^\star r^2 dr = 
L_\star \int \kappa_{\rm F} \rho dr 
 \equiv L_\star \tau_{\rm F} \ ,
\label{tflux}
\end{equation}
where $\tau_{\rm F}$ is the {\it flux mean optical depth}. From eqs.~(\ref{trep}) and (\ref{tflux}) we see that {\it in the optically thin limit, $\tau_{\rm rep}$ is equivalent to $\tau_{\rm F}$}.

For larger opacities, $I_{\lambda}^{\rm{env}}$ of eq.~\ref{lenv} is not zero, and its determination must take into account the full radiative transfer problem, including multiple scattering effects. In general, multiple scattering will increase the effective opacity of the envelope, making $\tau_{\rm rep}$ larger than $\tau_{\rm F}$. 
From this it follows that the detailed shape of the scattering phase function (SPF) can be of importance for determining the amount of reprocessing in the wind. 
That is the main reason why the SPF is listed above (item v in section~\ref{introduction}) as one of the quantities necessary to completely specify the radiative transfer problem in dusty media. In section~\ref{SPF} we will expand on this issue with further details.

An important consequence of eq.~(\ref{lenv}) it that, because $L_{\rm rep}$ depends on the details of the radiative transfer, it is not straightforward to obtain a relation between $L_{\rm rep}$ (or $\tau_{\rm rep}$) and the optical depth at the fiducial wavelength (say, $\tau_{V}$). In this paper we demonstrate that, because of AI, $\tau_{\rm rep}$ is the most suitable parameter for SED classification; it follows, then, that the optical depth $\tau_{V}$
is {\it not} a suitable parameter, contrary to the usual approach in the literature.

Although $\tau_{\rm rep}$ (or $L_{\rm rep}$) is a well defined quantity, relating this quantity to the physical parameters of the model  (such as mass loss rate) is a rather involved problem, requiring the full solution of the radiative transfer. On the other hand, in many instances $L_{\rm rep}$ is a well defined observational quantity. This is the case, for example, for optically thin envelopes surrounding a hot star, in which the attenuated stellar spectrum can be easily separated from the IR emission of the grains.
In the following, where we discuss the detailed consequences of AI, we will show that this issue is 
at the heart of the fundamental uncertainties in determining mass loss rates using the SED.

Finally, condition 3 of AI determines the radius of the cavity. In particular, the radius will be the dust destruction radius when the presence of the cavity is controlled by dust sublimation/condensation.
Note that by specifying the cavity radius in stellar radii we are not introducing another radial scale in the problem.

\section{The Monte Carlo Code \label{mc}}

For the calculations shown in this paper, we have used a Monte Carlo code developed to solve the general problem of the radiative transfer plus radiative equilibrium in dusty media \citep{car01}. The code uses a standard Monte Carlo simulation, which basically follows the path of a large number of {\it photon packets} as they are scattered, absorbed and reemited within a prescribed medium (for further details see, e.g., \citet{cod95}). The radiative equilibrium is solved using the method described in \citet{bjo01}, in which the grain temperature and emitted spectrum are corrected in the course of the simulation.

Although the radiative equilibrium calculation does not require any iteration, quantities like the dust condensation radius $r_{\rm i}$ and the reprocessing optical depth $\tau_{\rm rep}$ (see section~\ref{as}) do require iteration, since they depend on the results of the radiation transfer. Each iteration returns a temperature of the inner cavity and the amount of reprocessed energy. The values of $r_{\rm i}$ and the optical depth are modified accordingly, and a new simulation is run, until the correct values are obtained.

The code is fully 3-D, so it is capable of solving the radiative transfer problem for many circumstellar geometries and density distributions, including the presence of multiple sources. The transfer of polarized radiation is included, so the code is capable of providing the entire SED, the polarization as a function of wavelength, images at specific wavelengths as well as polarization maps.

The code treats an arbitrary mixture of grain species (amorphous carbon, silicate, etc.) and a different size distribution can be assigned to each grain species. The grain scattering and absorption properties are calculated using Mie theory, without approximations, to obtain the correct form of the scattering phase function for polarized incident radiation. The radiative equilibrium condition is imposed separately for each grain type and size, so the code calculates independent radiative equilibrium temperatures for each grain type and size. 
This procedure contrasts with the usual procedure in the literature, which consists of assigning a single temperature to all grain sizes, whose opacity and scattering properties are obtained by averaging over the size distribution.
In section 5, we will show that, for hot stars mainly, individual grain sizes can have very different equilibrium temperatures, which has important effects for the model IR SED. 

In summary, the only approximation used in our dust model is that the shape of dust grain is spherical. 

\section{Single Grain Size Models \label{single}}

In this section, we explore the effects of grain size on the SED for single grain size models, and interpret the results in the framework of approximate invariance. 
Our goal is to determine to what extent the grain size can be varied while maintaining the approximate invariance of the SED.
We begin by defining the basic model we use in this and in the next section. It consists of a central star
of unit radius
 that emits a black body spectrum of temperature $T_{\rm eff}$, surrounded by a spherical dust shell with internal radius $r_{\rm i}$ and external radius $1000\, r_{\rm i}$. 
 The internal radius is set by the temperature of the cavity wall, $T_{\rm i}$, often taken to be the dust condensation temperature.
The dust density profile is proportional to $r^{-2}$, and the dust grains are spherical. This is a simple model, but useful with respect to many astrophysical situations, such as the nearly spherical circumstellar envelopes found around many RGB and AGB stars.

Three different chemical compositions were studied: cosmic silicate, with optical constants given by \citet{oss92}, amorphous carbon (optical constants by \citet{zub96}) and silicon carbide (optical constants by \citet{peg88}). The sublimation temperature for these materials was taken to be $1000\,\rm K$ for silicate, $800\,\rm K$ for amorphous carbon and $1500\,\rm K$ for silicon carbide. These numbers are somewhat arbitrary, but they reflect approximately the temperatures used in the literature (e.g., IE97, and \citet{lor94}).

The model parameters we vary are the dust properties, optical depth, source temperature, and temperature of the cavity wall. To comply with the three conditions of AI, 
stated in section~\ref{as}, 
we compare models with the same chemical composition, the same temperature of the cavity wall, and the same reprocessing optical depth. 

We shall distinguish between three different cases, corresponding to three optical depth regimes, which will be studied separately. In case A, the envelope is optically thin, both in the wavelengths emitted by the source (the extinction region) and in the IR (the reprocessing region). In case B, the envelope is optically thick in the extinction region, but is optically thin to reprocessed radiation, and in case C the envelope is optically thick in both regions (up to the wavelength for which the Rayleigh limit is achieved for all grain sizes considered).

\subsection{Case A: Optically Thin Envelopes \label{caseas}}

In this case, we determine the V-band optical depth $\tau_{V}$ for each model using the condition $\tau_{\rm rep}=0.1$, which corresponds to reprocessing 9.5\% of the input radiation. In the extinction region, about 90\% of the flux is stellar in origin, so we expect SEDs from different models will be similar in the extinction region, at least at the 10\% level. The IR SED will also be similar, as a result of the AI. Hence, we expect to observe a similarity between all models for all wavelengths.

We illustrate this situation in Figure~\ref{sedas}. Each panel shows the SED for four different models, each with a different grain size, for cosmic silicate grains with a given stellar temperature. In Figure~\ref{tauas}, we show the extinction optical depth as a function of wavelength for the models with $T_{\rm eff}=2500\,\rm K$. The other model parameters, cavity radius, grain mass and V-band optical depth are listed in Table~\ref{tab1}.

Figure~\ref{sedas} corroborates our above qualitative discussion. For all grain sizes and stellar temperatures, the short wavelength side of the curve is very similar, as a consequence of the low optical depth combined with the same total attenuation. Similarly, the overall amplitude of the IR emission scales approximately with $\tau_{\rm rep}$ and the shape is set by the grain absorption efficiency, which is the same for each grain size. As a result, the overall SED is very similar for all models. The mean difference between the models with $T_{\rm eff}=2500\, \rm K$ is about 7\%, and for the models with $T_{\rm eff}=20000\,\rm K$, the mean difference is 9\%. The similarity between the model SEDs, not only in the IR but also in the extinction region, allows us to extend the consequence of AI. We conclude that in the optically thin limit {\it the entire SED is similar when the conditions of AI are met}.

The conditions for approximate invariance are violated when we consider grains larger than about $a = 0.25 \,\mu \rm m$. These grain sizes fail to satisfy condition 1 of AI, so large differences in the IR SED of these models are observed. This is shown in Figure~\ref{grande}, where we plot the SEDs for models with grain sizes $0.005, 0.50$ and $1.0\,\mu \rm m$. The large differences in the IR SED are apparent. 
However, as shown in eq.~(\ref{eq1}), the largest grain that satisfies condition 2 of AI depends on the temperature of the cavity wall. Consequently, if we lower the condensation temperature, these larger grains will also satisfy the AI conditions. This is illustrated in Figure~\ref{sed300}, which shows the SED for envelope models with $T_{\rm i} = 300\,\rm K$. These plots show that a distinction can hardly be made between envelope models with grain sizes now ranging up to $1 \,\mu \rm m$. For $T_{\rm eff}=2500 \rm K$ the mean difference between the SEDs is at most 3\%. 
This has the consequence that the SED is a poor constraint for grain sizes, at least for sizes that satisfy the conditions of AI. 

We conclude that, for case A, given a grain composition, a temperature for the inner cavity, and a spatial distribution of the dust, the SED depends on a single parameter, $\tau_{\rm rep}$. 
Another important result is that $\tau_{\rm rep}$ is not uniquely related to the  $V$-band optical depth (or any other wavelength), as shown in Figure~\ref{tauas}. All curves are for $\tau_{\rm rep}=0.1$, and yet show very large differences  in the optical depth for most wavelengths. 
From this we can conclude that {\it the optical depth is not a suitable parameter for SED classification}; if we were to group the case A models according to $\tau_{V}$, for example, the models would have large differences in the IR SED, leading us to the wrong conclusion that the grain size does have an important effect on the SED, which we have demonstrated not to be the case.

The optical depth depends on the product of the grain opacity (which depends on the grain size) and the grain density.
The fact that grain sizes are very difficult to determine from the SED implies a corresponding uncertainty in the dust mass of the envelope, as shown in table~\ref{tab1}. This uncertainty directly affects the ability to measure the mass loss rates from stars using the SED.

\subsection{Case B: Optically Thick Extinction with Optically Thin IR Emission \label{casebs}}

In this case, the envelope is optically thick in the UV and visible, and optically thin in the IR. Based on AI, we expect the behavior of the IR SED to be similar to case A; i.e., the shape and level of the IR emission should be the same for all models with the same $\tau_{\rm rep}$.
In contrast to case A, the larger optical depths result in a much larger attenuation of the stellar flux. Therefore, the shape of the SED in the extinction region is controlled by the shape of the extinction optical depth, which strongly depends on grain size.

\begin{figure}[bp]
\plottwo{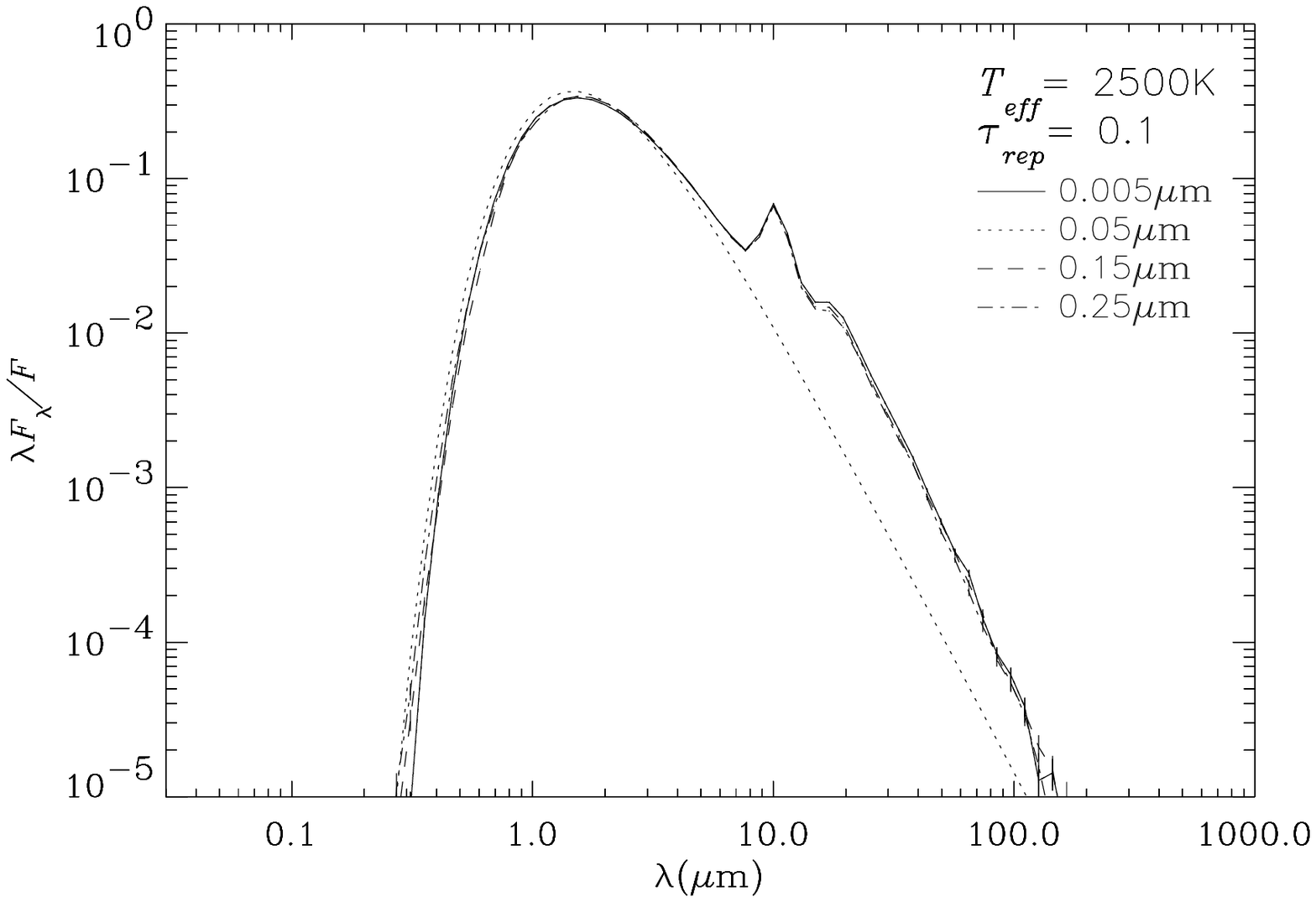}{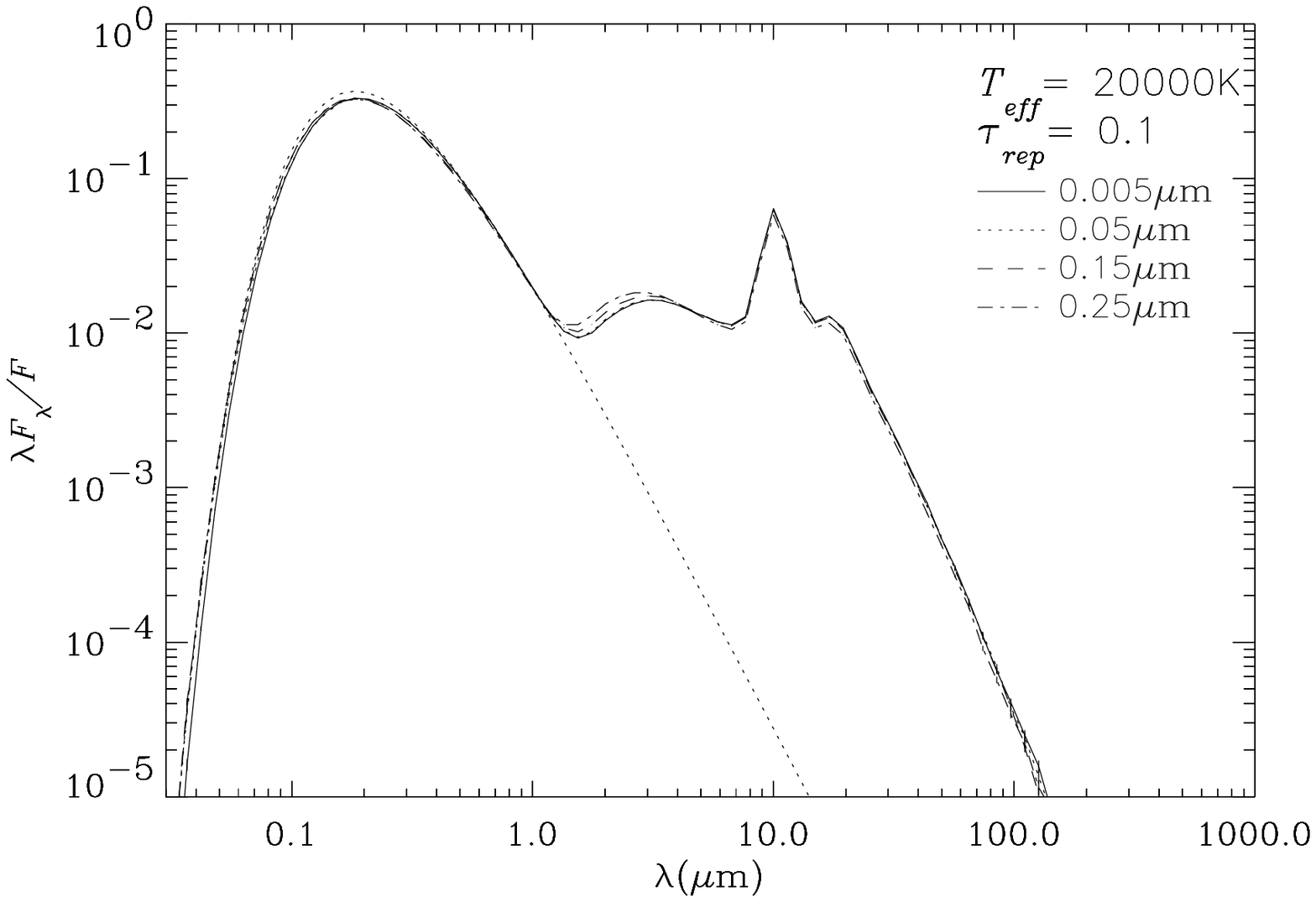}
\caption[]
{Single grain size models for case A, with $T_{\rm i} = 1000\,\rm K$. Shown are the results for four envelope models of cosmic silicate grains, with $a = 0.005, 0.05, 0.15$ and $0.25 \,\mu \rm m$. The left and right panels show the results for a given stellar temperature, as indicated. For each model, the optical depth was adjusted so that $\tau_{\rm rep}=0.1$ (see Table~\ref{tab1}). Also shown is the stellar spectrum ({\it dotted line}). \label{sedas}}
\end{figure}

\begin{figure}[bp]
\plotone{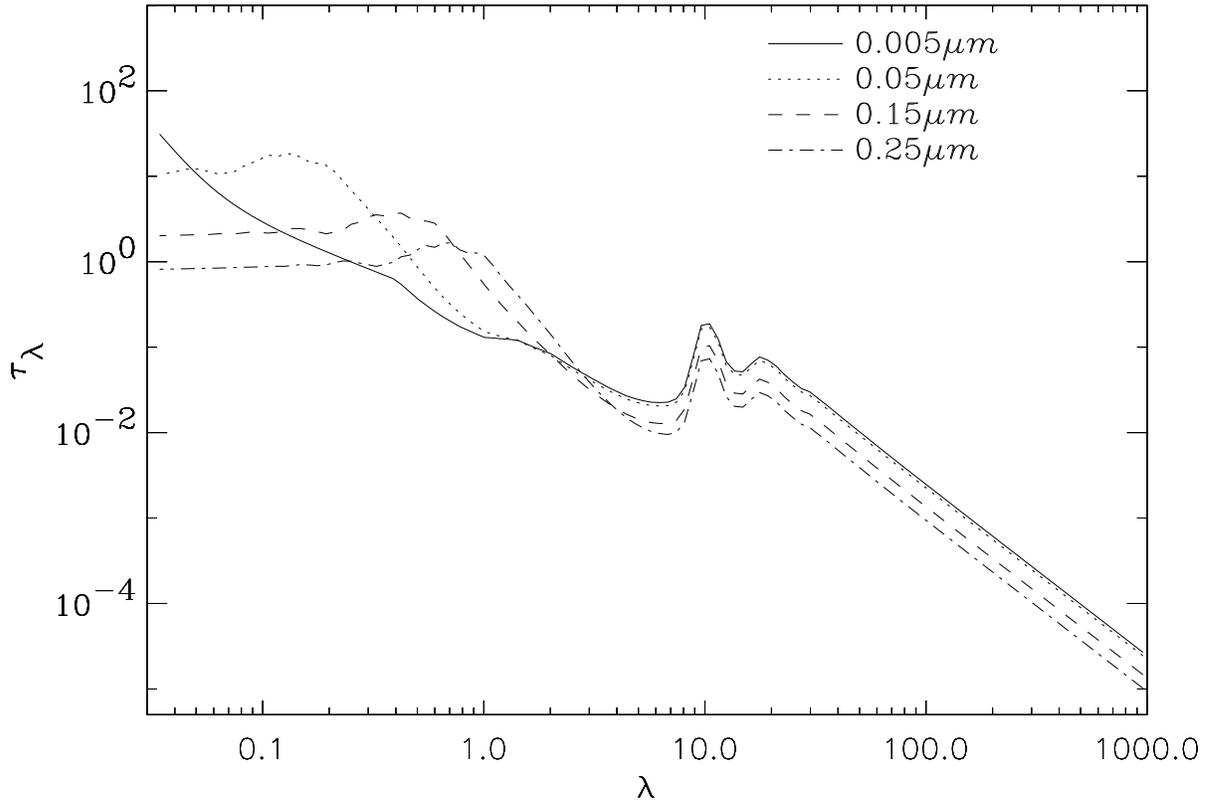}
\caption[]
{Extinction optical depth. The optical depths $\tau_\lambda$ vs. wavelength is shown for the models of Figure~\ref{sedas}, for the case $T_{\rm eff} = 2500\,\rm K$. Note that $\tau_\lambda \lesssim 1$ for the wavelengths emitted near the peak of the source flux ($1\,\mu \rm m < \lambda < 3\,\mu \rm m$), as required for case A. \label{tauas}}
\end{figure}

\begin{figure}[bp]
\plottwo{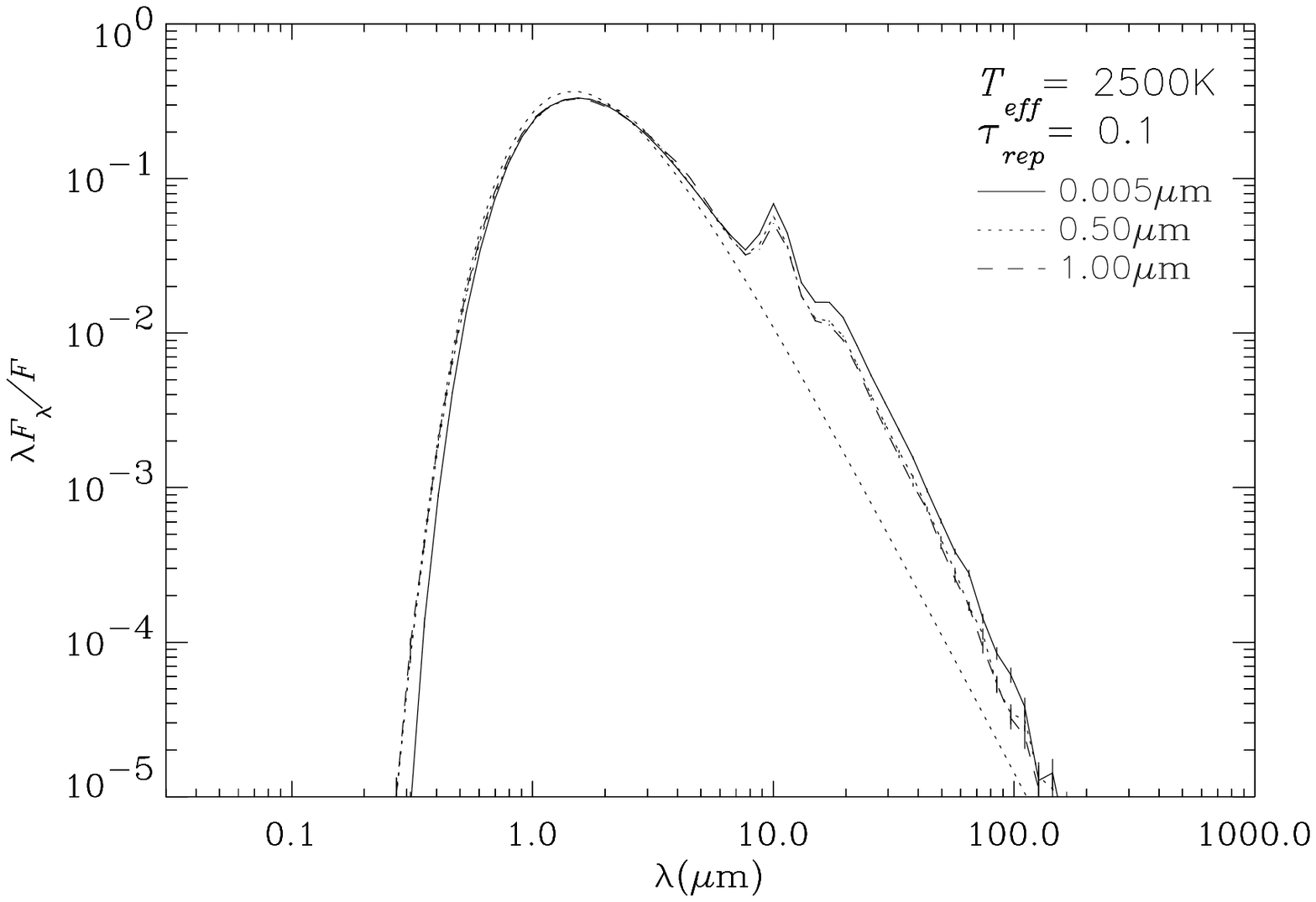}{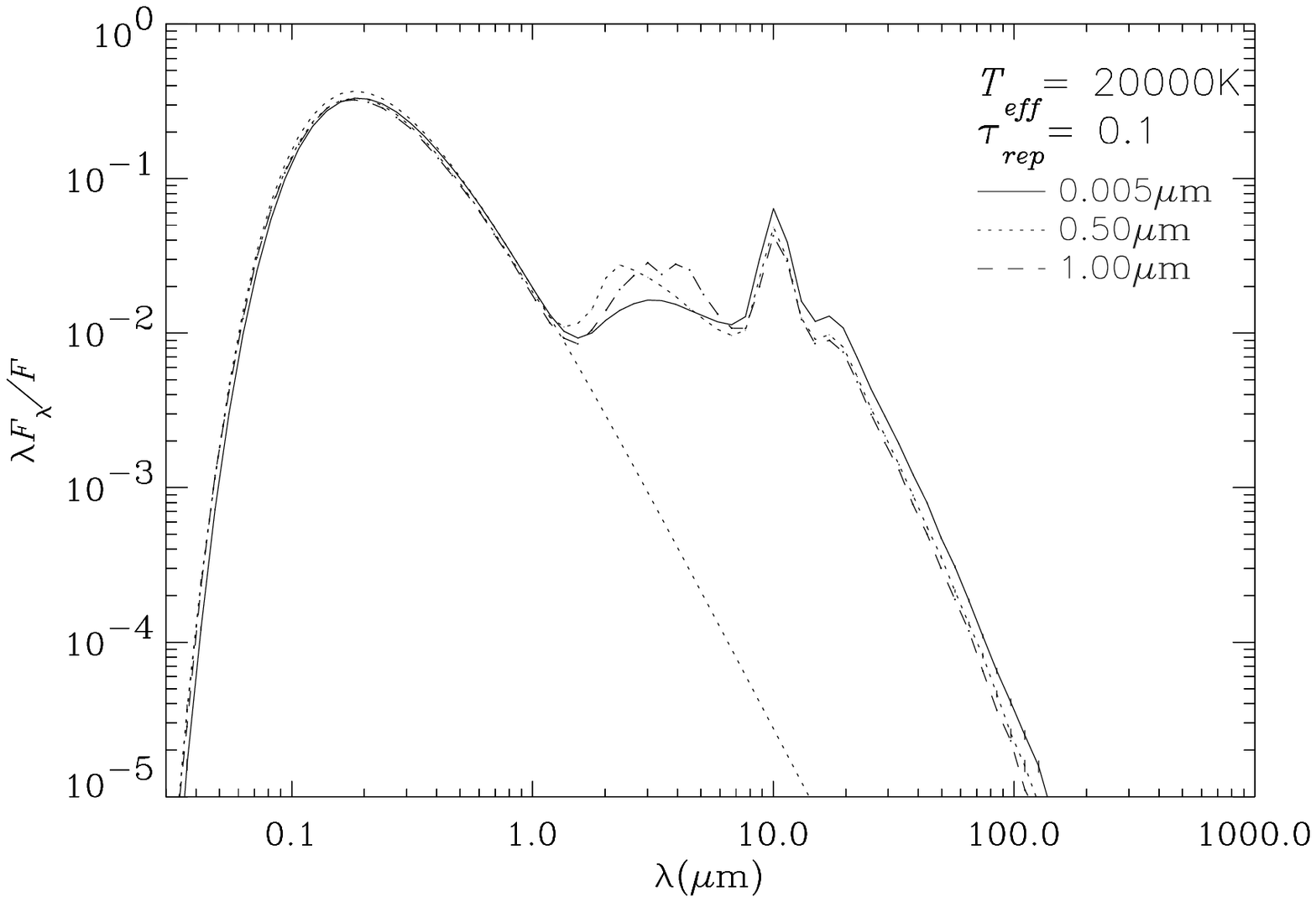}
\caption[]
{Single grain size models for case A, with $T_{\rm i} = 1000\,\rm K$. Shown are the results for three envelope models of cosmic silicate grains, with $a = 0.005, 0.50$ and $1.0 \,\mu \rm m$. The left and right panels show the results for a given stellar temperature, as indicated. For each model, the optical depth was adjusted so that $\tau_{\rm rep}=0.1$. Also shown is the stellar spectrum ({\it dotted line}). \label{grande}}
\end{figure}

\begin{figure}[bp]
\plottwo{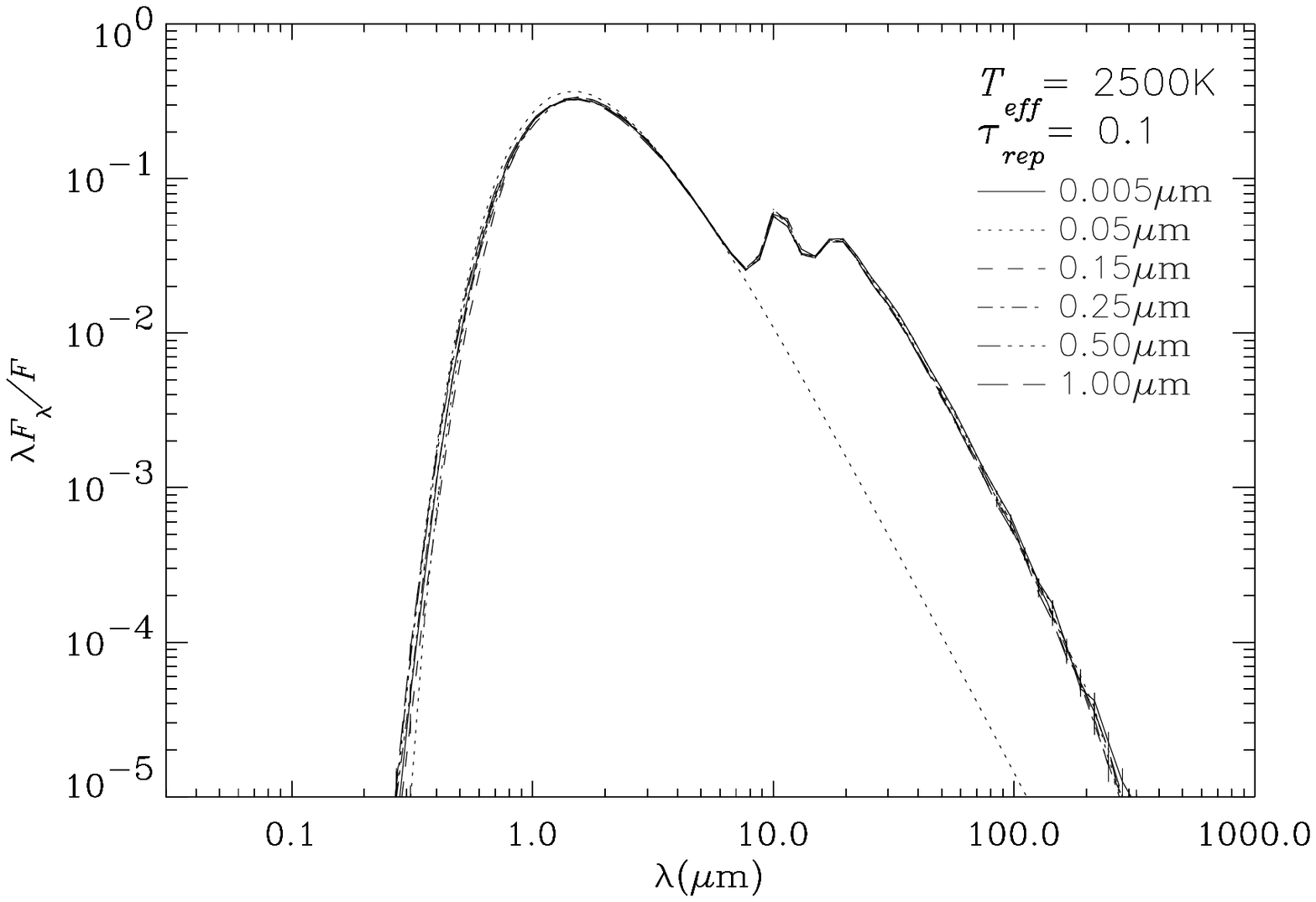}{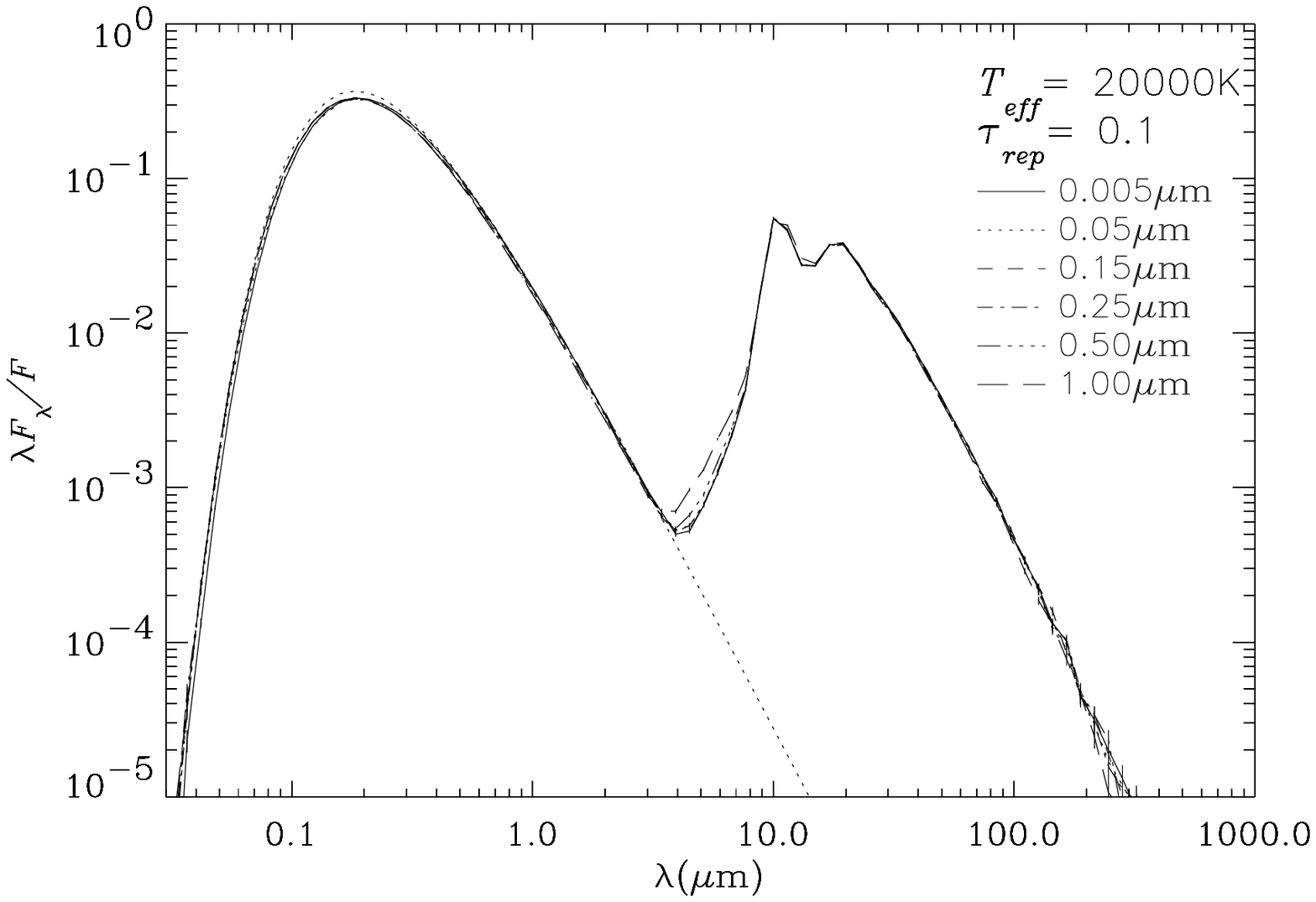}
\caption[]
{Single grain size models for case A, with $T_{\rm i} = 300\,\rm K$. Shown are the results for six envelope models of cosmic silicate grains, with $a = 0.005, 0.05, 0.15, 0.25, 0.50$ and $1 \,\mu \rm m$. The left and right panels show the results for a given stellar temperature, as indicated. For each model, the optical depth was adjusted so that $\tau_{\rm rep}=0.1$. Also shown is the stellar spectrum ({\it dotted line}). \label{sed300}}
\end{figure}

\begin{figure}[bp]
\plottwo{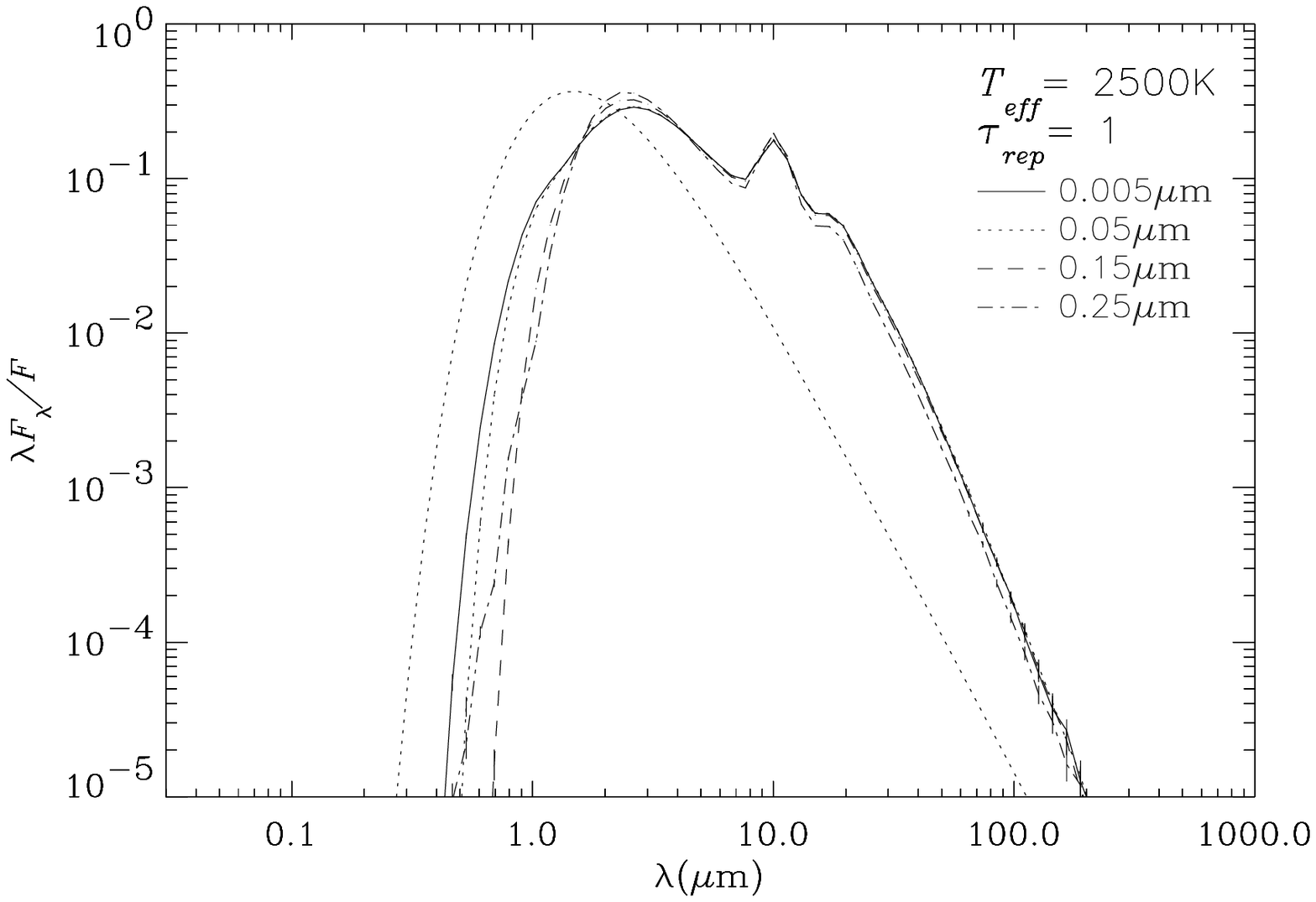}{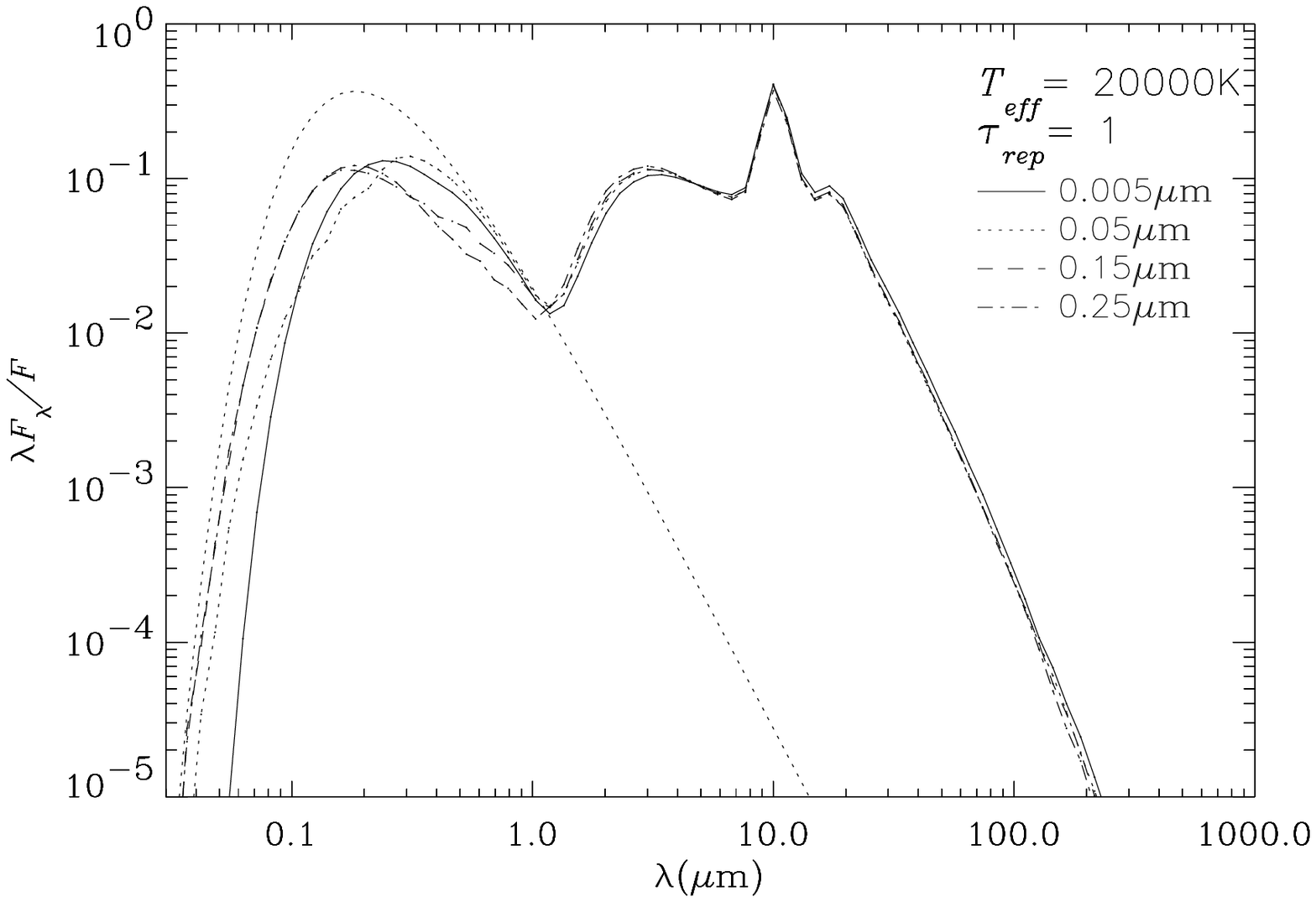}
\caption[]
{Single grain size models for case B, with $T_{\rm i} = 1000\,\rm K$. Shown are the results for four envelope models of cosmic silicate grains, with $a = 0.005, 0.05, 0.15, 0.25 \,\mu \rm m$. The left and right panels show the results for a given stellar temperature, as indicated. For each model, the optical depth was adjusted so that $\tau_{\rm rep}=1$. Also shown is the stellar spectrum ({\it dotted line}). \label{sedbs}}
\end{figure}

\begin{figure}[bp]
\plotone{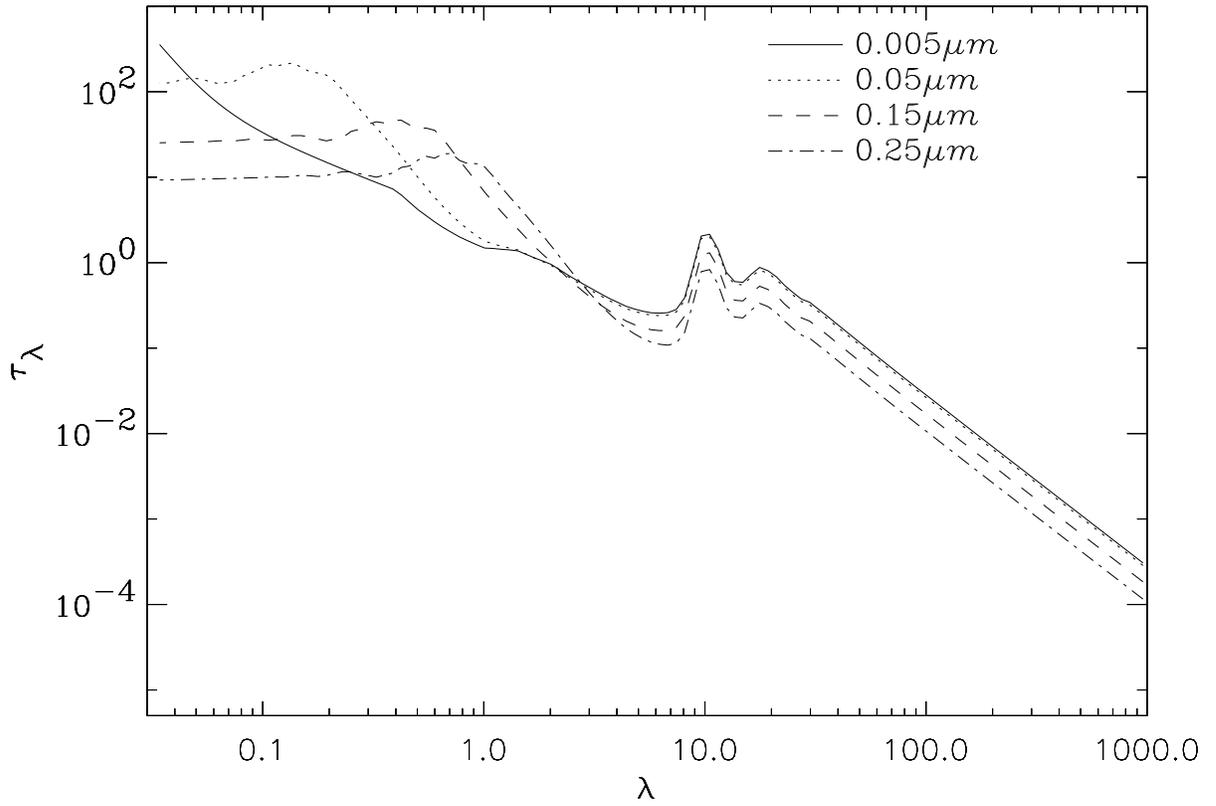}
\caption[]
{Extinction optical depth. The optical depth $\tau_\lambda$ vs. wavelength is shown for the models of Figure~\ref{sedbs}, for the case with stellar temperature of $2500\,\rm K$. \label{taubs}}
\end{figure}

In Figure~\ref{sedbs} we present the SED for the same envelope models as in Figure~\ref{sedas}, but we maintain $\tau_{\rm rep}=1$. This implies 63.2\% of the input radiation is reprocessed by the envelope. As before, other models parameters are listed in Table~\ref{tab2}, and in Figure~\ref{taubs} we display the extinction optical depth as a function of wavelength.

We see in Figure~\ref{sedbs} that indeed the IR SED for the different models are similar, while the SEDs for short wavelengths are very different because they depend on the shape of the grain efficiencies. It follows that case B is different from case A, in the sense that the SED does depend strongly on the dust grain size. There are, however, two points to be considered. First, in order to obtain information about grain size, one has to observe the SED in the wavelengths where the effects of grain sizes are important (i.e., UV, visible or NIR, depending on the model). Second, single grain size models are very crude approximations for actual stellar envelopes, where a distribution of grain sizes is expected. In section~\ref{dist} we study this case and show how it affects our conclusions for case B.

At this point it is useful to compare our results with those of previous authors. 
For example, IE97 show SEDs in their Figure 10 for spherical envelopes of silicate and amorphous carbon dust grains, of varying sizes and optical depths. 
In their plots on the right, where the results for silicate grains are displayed, the upper two plots can be compared roughly with our Figures~\ref{sedas} and~\ref{sedbs}.
In Figure 10 of IE97, large differences between the models with $a = 0.05$ and $0.1 \mu \rm m$ are evident. By looking at these plots, the reader is led to believe that the IR SED for grains with sizes $0.05$ and $0.1 \mu \rm m$ is intrinsically different. However, our results show that this is not the case. The IR SED of single grain size models whose sizes comply with the conditions of AI have very similar shapes and levels.

The differences between Figure 10 of IE97 and our Figures~\ref{sedas} and~\ref{sedbs} owe to the choice of the model optical depth. In IE97, all models have the same extinction  $V$-band optical depth. Our models, on the contrary, have all the same $\tau_{\rm rep}$ but very different $\tau_{V}$, as seen in Figures~\ref{tauas} and~\ref{taubs}.
Thus in case B, we also find that $\tau_{\rm rep}$ is the best parameter to reveal the essential similarities of models with different grain sizes (which must be lower than the limit set by condition 2 of AI).
\rm

\subsection{Case C: Optically Thick Envelopes also in the IR \label{casecs}}

In this case, the envelope is so optically thick that 100\% of the input radiation is reprocessed by the dust grains. As a consequence, the quantity $\tau_{\rm rep}$ becomes ill-defined, so another parameter must be found to reveal the similarities resulting from AI.

By definition, all models, in this case, are optically thick to the reprocessed radiation. This means that most of the radiation will emerge at wavelengths larger than the Wien peak of the cavity wall (i.e., $\lambda > 3 \,\mu \rm m$ for $T_{\rm i}=1000 \rm K$). 
For a given wavelength the envelope can be divided into two different regions with different properties concerning the radiation field: an inner region with radius between $r_{\rm i}$ and $r_{\tau=1}$, and an outer region, between $r_{\tau=1}$ and $r_{\rm e}$, where $r_{\rm e}$ is the envelope outer radius, and $r_{\tau=1}$ is given by
\begin{equation}
\int_{r_{\tau=1}}^{r_{\rm e}}\rho \, \kappa_{\lambda} dr = 1 \ , \label{tauc}
\end{equation}
i.e., $r_{\tau=1}$ is the point where the radial optical depth of the envelope is unity (this quantity depends on the wavelength). By definition, radiation emitted  inside the inner region is unlikely to leave the envelope because the optical depth is larger than 1; hence, a reasonable approximation for the emerging IR flux is given by the volume integral of the radiation emitted by the outer (optically thin) region
\begin{equation}
F_\lambda = \int_{V} 4\pi j_\lambda dV = 16\pi^2 \int_{r_{\tau=1}}^{r_{\rm e}} r^2 \rho(r)\, \kappa_\lambda B_\lambda[T(r)] dr \ . \label{emis1}
\end{equation}

In order to find an analytic expression for $F_\lambda$, we will suppose that both the temperature in the outer part of the envelope and the grain opacity can be approximated by power-laws, so we write
\begin{equation}
T(r)=T_0 \left({r_{\rm i} \over r}\right)^s \ , \label{temp}
\end{equation}
and
\begin{equation}
\kappa_\lambda=\kappa_0 \left({\lambda_0 \over \lambda}\right)^p \ , \label{kappa}
\end{equation}
where $\lambda_0$ is an arbitrary constant. The assumption for the opacity is very reasonable, because by definition most of the flux emerges at wavelengths larger than the Rayleigh limit (see Figure~\ref{effsil} and discussion in section~\ref{as}). The assumption for the temperature is justified by the fact that the outer part of the envelope is optically thin; it can be shown that, if the dust opacity is given by eq.~(\ref{kappa}), the radial dependence of the temperature of an optically thin envelope is $T(r) \propto r^{-2/(p+4)}$ (e.g. \citet{mar78}, eq. 7.3). Our detailed calculations, shown below, validate this assumption.
Finally, assuming a dust density $\rho(r)$, also given by a power-law
\begin{equation}
\rho(r)=\rho_0\left({r_{\rm i} \over r}\right)^n \ , \label{dens}
\end{equation}
we calculate $r_{\tau=1}$ using eqs.~(\ref{tauc}), (\ref{kappa}) and (\ref{dens}), which gives
\begin{equation}
\frac{r_{\tau=1}}{r_{\rm i}}=\left(\frac{\lambda_0}{\lambda}\right)^{\frac{p}{n-1}} \tau_0^{\frac{1}{n-1}} \ , \label{r1}
\end{equation}
where
\begin{equation}
\tau_0 \equiv \frac{\kappa_0\rho_0 r_{\rm i}}{n-1} 
\end{equation}
is the envelope optical depth at $\lambda = \lambda_0$.

Substituting eqs. (\ref{temp}), (\ref{kappa}), (\ref{dens}), and {(\ref{r1})} in eq.~(\ref{emis1}) and letting $r_{\rm e} \rightarrow \infty$, we obtain the spectral shape of the radiation emitted by the outer part of the envelope
\begin{equation}
{\lambda F_\lambda \over L_\star} = K \lambda^{\frac{3-n}{s}-p-5} \int_{u_1}^{\infty} {u^{2-n} \over e^{u^s}-1}du \ , \label{flux}
\end{equation}
where
\begin{equation}
u_1 = \Gamma \left( \frac{1}{\lambda} \right)^{\frac{n-1+sp}{s(n-1)}} \ ,
\end{equation}
and 
\begin{equation}
\Gamma \equiv \frac{\left({k T_0 \over h c}\right)^{s}}{\left(\tau_0 \lambda_0^p \right)^{n-1}} \ . \label{gamma}
\end{equation}
The normalization constant $K$ is defined by the condition
\begin{equation}
\frac{\int_{0}^{\infty} \lambda F_\lambda d\lambda}{L_\star} = 1 \ .
\end{equation}

In eq. (\ref{flux}) the only parameter that depends on the grain size is $\Gamma$ in the lower integration limit, $u_1$. It follows that, if the values of $\tau_0$ and $T_0$ of models with different grain sizes are such that $\Gamma$ is the same, {\it these models will produce the same SED}. 

\begin{figure}[bp]
\plotone{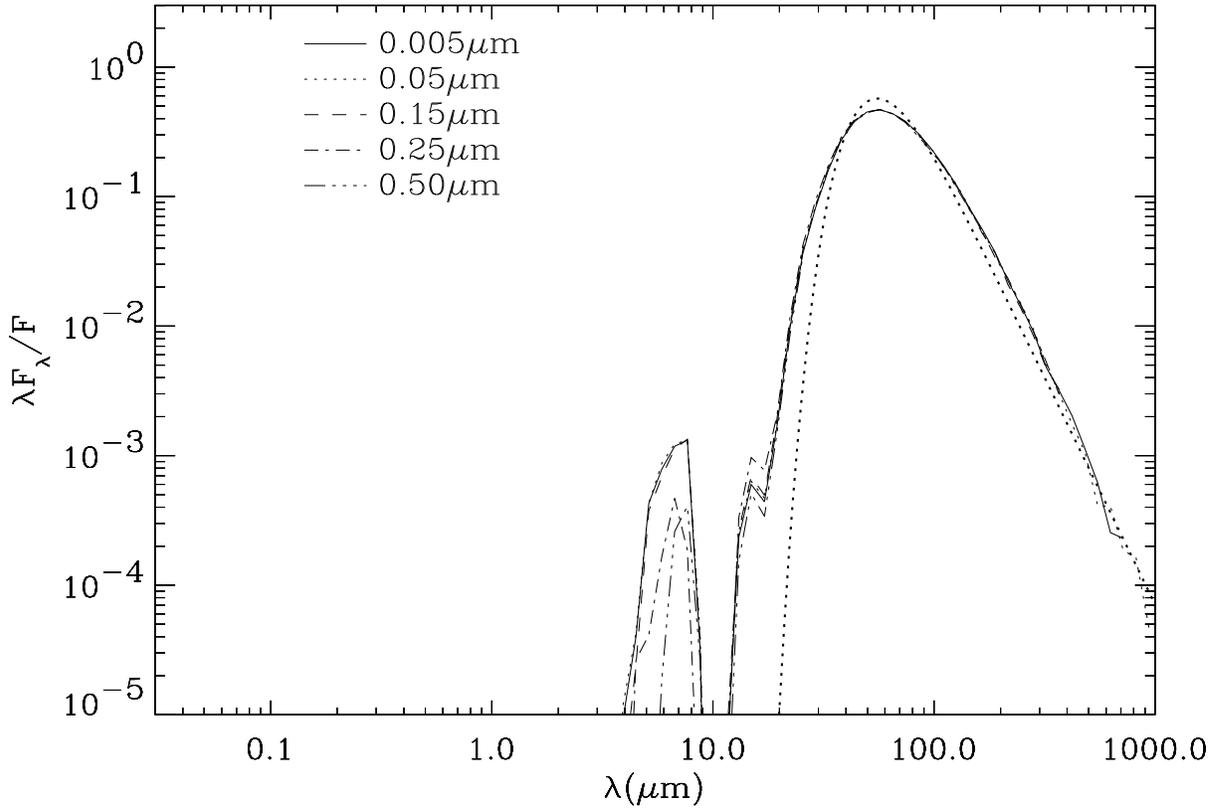}
\caption{SEDs for five envelope models of cosmic silicate grains, with $a = 0.005, 0.05, 0.15, 0.25$ and $0.5 \,\mu \rm m$. The stellar temperature is $20000\,\rm K$. Model optical depths are given in Table~\ref{tab3}. The temperature of the inner cavity was set to $1000\,\rm K$. Also shown is the curve for eq.~(\ref{flux}) ({\it thick dotted line}). \label{sedcs2}}
\end{figure}

\begin{figure}[bp]
\plotone{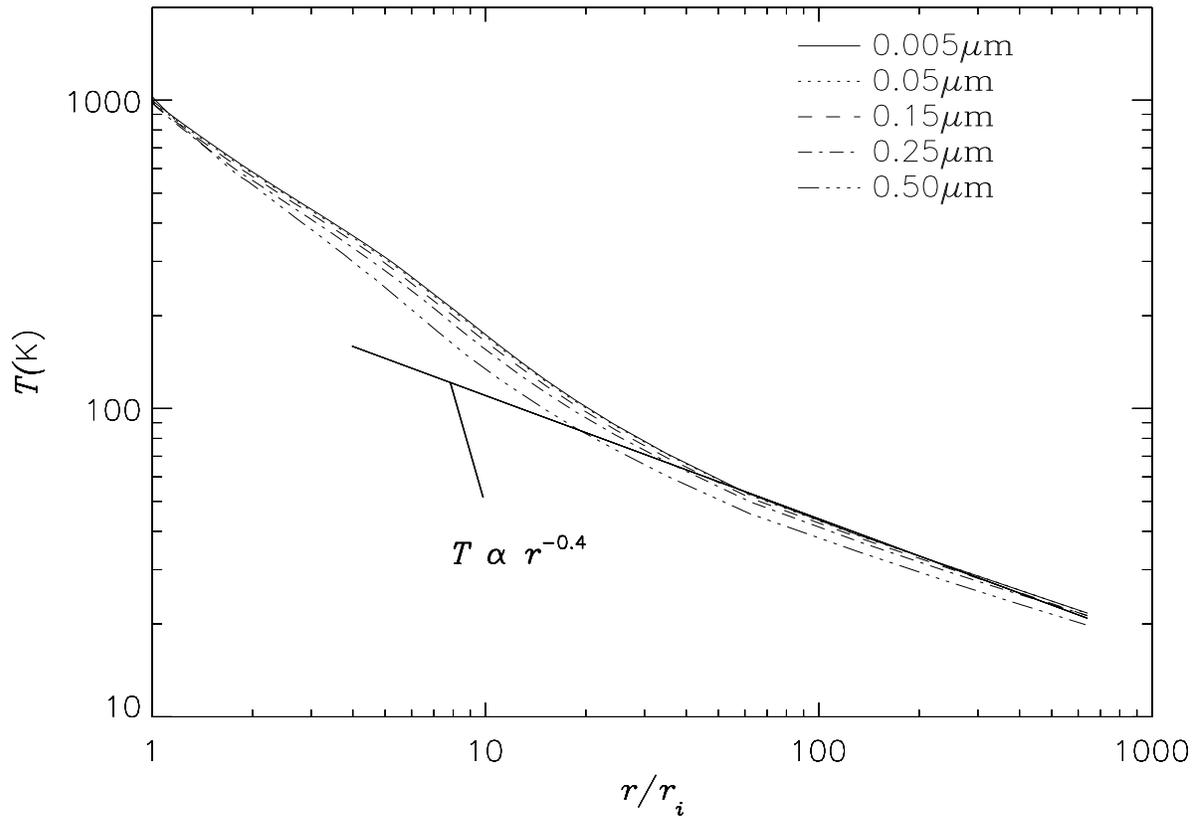}
\caption{Temperature profiles for the same models of Fig.~\ref{sedcs2}. Also shown is a fit for the temperature of the model with $a=0.05 \,\mu \rm m$ in the outer part of the envelope, with a power-law of index $s=0.4$. \label{tempc}}
\end{figure}

We show in Figures~\ref{sedcs2} and~\ref{tempc} the SEDs and grain equilibrium temperatures for $20000\,\rm K$ models with different grain sizes. 
Figure \ref{tempc} shows that the temperature power-law exponent has a value $s=0.4$ at large $r$.
Other model parameters are listed in Table~\ref{tab3}. Note that the optical depths are very high (ranging from $\tau_0 = 3.4$ to 5.0 for $\lambda_0=100\,\mu\rm m$), according to the assumption for case C. These optical depths were chosen so that the parameter $\Gamma$ is the same among the different models (i.e., we measured $T_0$ and adjusted $\tau_0$ so that $T_0^s \tau_0^{1-n}=$ constant). 
The resulting SEDs are very similar for $\lambda\gtrsim 20 \mu\rm{m}$, which indicates that the parameter $\Gamma$ is an adequate scaling parameter for very high optical depths. 
For shorter wavelengths the SED invariance breaks down, because the opacities are no longer described by a power-law, and the SEDs are different.

Also shown in Figure~\ref{sedcs2} is the analytic expression for the emerging flux, eq.~(\ref{flux}). This expression reproduces well the shape of the SED for long wavelengths; for shorter wavelengths the agreement is not good for the reason stated above: our assumption for the grain opacity fails in the well-known silicate spectral features at 10 and 20 $\mu\rm m$, and as we go into the optical limit.

We conclude that, given a dust grain composition and spatial distribution, the shape of the IR SED is controlled by a single parameter, $\Gamma$.  
From the observational point of view, we have a situation similar to case B, where the IR emission bump does not convey any information about the dust grain size.
However at shorter wavelengths, AI breaks down, so one can determine information about the dust grain size, if there is sufficient observable flux at these wavelengths.

\subsection{Effects of the Scattering Phase Function on the SED \label{SPF}}

In the previous sections we studied in detail the effects of grain size on the SED, and determined to what extent the grain size (i.e., opacity) can be changed while maintaining the approximate invariance of the SED.
An additional effect is that, when the grain sizes are changed, both the opacity and the SPF change. 
 Changing the SPF breaks invariance requirement (v) of section~\ref{introduction} and thus will alter the SED.
The results shown in the previous sections were calculated using Mie theory which has an anisotropic SPF, but a common simplification is to use an isotropic SPF. In this section, we briefly discuss how an isotropic SPF affects our previous results.

In general, the primary change is the $V$-band optical depth required to reproduce $\tau_{\rm rep}$; the SED is largely unaffected (except at the few percent level).
In Table~\ref{tab4}, we compare the  $V$-band optical depth between models with an anisotropic SPF ($\tau_{V}$) and an isotropic SPF ($\tau_{V}^{\rm iso}$) for both case A and B. 
One should compare column 1 with column 2, and column 4 with column 5.
Because all models are subject to the condition $\tau_{\rm rep} = 0.1$ (case A) or 1 (case B), the difference between  $\tau_{V}$ and $\tau_{V}^{\rm iso}$ for each case gauges the effects of the SPF on the amount of reprocessing for each model.

For case A models, $\tau_{V}$ and $\tau_{V}^{\rm iso}$ are roughly the same (at the 10\% level). This is to be expected because, in the optically thin limit, $I_\lambda^{\rm env}$ in eq.~(\ref{lenv}) is small, so the SPF ($d\sigma/d\Omega$) has little effect on $L_{\rm rep}$.

For case B models, the differences between $\tau_{V}$ and $\tau_{V}^{\rm iso}$ are larger, but still not very large (at most 22\%). We conclude that for optically thin models the SPF is largely irrelevant, while for optically thick models, the SPF affects the conversion from $\tau_{V}$ to $\tau_{\rm rep}$, but at most to a level at 20\% to 30\%.

Also shown in Table~\ref{tab4} is the flux mean optical depth, $\tau_{\rm F}$, defined in eq.~\ref{tflux}. 
One should compare these values with the corresponding value of $\tau_{\rm rep}$. 
The difference between  $\tau_{\rm F}$ and $\tau_{\rm rep}$ arises from multiple scattering effects, and for this reason they are much smaller for optically thin models than for optically thick models.
Many authors use $\tau_{\rm F}$ as a parameter for SED classification; the results in Table~\ref{tab4} show that this is not a good parameter, at least for optical depths in the range of case B models, for which multiple scattering effects are important.

\section{Models with a Grain Size Distribution \label{dist}}

The single grain size assumption can be a poor approximation for describing the dust of an astrophysical system. For example, fitting the interstellar extinction curve requires a distribution of grain sizes. \citet[hereafter MRN]{mat77} used a simple power law size distribution with index $q=-3.5$, and a lower and upper cutoff size, $a_{\rm min} \sim 0.05\,\mu\rm m$ and $a_{\rm max} \sim 0.25\,\mu\rm m$. This distribution has been revised many times in the literature. For example, \citet{kim94} proposed a distribution in the form
\begin{equation}
\frac{dn}{da}\propto a^{-q}e^{-a/a_0}, a \geq a_{\rm min} \ , \label{eq2}
\end{equation}
where $a_0=0.14\,\mu\rm m$ for silicate grains and $a_0=0.28\,\mu\rm m$ for graphite grains. The upper cutoff was replaced by an exponential decay, to allow for the presence of large dust grains ($a \gtrsim 0.2 \,\mu \rm m$).

Although the fact that the dust has a distribution of grain sizes is well established, the exact form of this distribution is still an open issue. 
As we will demonstrate below, part of this uncertainty is due to the AI of the problem, which  makes two models with similar size distributions virtually indistinguishable.
Another source of uncertainty lies on our yet incomplete knowledge of the details of grain formation and growth.
To complicate matters even further, the grain size distribution is likely to vary across the wind because different grain sizes have different drift velocities \citep{eli01}.

In this section, we study the SED of models with a grain size distribution. 
For simplicity, we adopt a power-law (MRN) grain size distribution function; although the quantitative details of the results shown below do depend on the adopted size distribution, the fundamental concepts discussed do not.
Previously, in section~\ref{single}, we used the concept of AI to reveal important similarities in the IR SED (and in the extinction SED for optically thin models) for single grain size models. In the following, we study what controls the spectral shape of the SED for models with a size distribution in an attempt to identify the same sort of similarities.

\subsection{Spectral Shape of the SED \label{sec51}}

Let us consider an envelope model, consisting of spherical dust grains of the same chemical composition and sizes distributed according to a given size distribution function, $dn/da$. The envelope IR SED, ${F}_{\rm IR}$, will be given by the sum of the contributions from each grain size
\begin{equation}
{F}_{\rm IR}(\lambda)= \int_{a_{\rm min}}^{a_{\rm max}} f(a) \mathcal{F}_{\rm IR}(\lambda,a) da \ ,
\end{equation}
where $f(a)$ is the wavelength integrated emission for each grain size,
a measure of how the absorbed energy is split between the different grain sizes, and $\mathcal{F}_{\rm IR}(\lambda,a)$ is the spectral shape of the spectrum emitted by the grains with radius $a$.

The simplest situation is when the spectrum emitted by all grains has approximately the same spectral shape (i.e., $\mathcal{F}_{\rm IR}(\lambda,a)$ is independent of $a$). 
In this case, the spectral shape of the IR SED of the envelope will be similar to the spectral shape of the individual grains, regardless of how the absorbed energy is split between the different grains. 
A somewhat similar situation occurs when a particular range of grain sizes, with a similar emission spectrum, completely dominates the emission. Here, even if the spectral shape of the other grain sizes is different, it does not affect the envelope IR SED. An important difference from the first situation is that, here, the manner in which the absorbed energy is distributed between the grain sizes is important.
For both these cases, we expect a close similarity to the corresponding single-sized grain model.
An opposite situation occurs when different grain sizes emit a different spectral shape with similar integrated emissions. In this case, the envelope IR SED can no longer be described using a single grain size model. 

What controls $\mathcal{F}_{\rm IR}(\lambda,a)$ is the spectral shape of the absorption efficiencies and the grain temperature. From the conditions of AI, it is evident that the grain sizes with similar emission spectra will be the ones that comply with conditions 1 and 3.
The first condition requires all grain sizes of the distribution must be smaller than a maximum size, given by eq.~(\ref{eq1}).
The third condition implies that all grain sizes must have similar equilibrium temperatures. However, owing to their different absorption efficiencies, different grain sizes can have very different equilibrium temperatures. This effect can be inferred from the second column of Table~\ref{tab1}: different grain sizes have different condensation radii, which indicates that, if these grains were to coexist at the same point in space, the grain sizes with larger condensation radii will be hotter than those with lower condensation radii. Recently, \citet{wol03} studied the condensation temperature of the individual grain sizes in a mixture of different grains sizes, both in 1-D and 2-D dust shells, and found that the temperature difference spans a range of up to $\approx 250$ K, although this value is highly dependent of the choice of the dust properties.

It is easy to show that, in the Monte Carlo radiative equilibrium scheme, the temperature $T(a)$ of the grains with radius $a$ at a given point of the envelope will depend on the number of photons absorbed by these grains, $N_{\rm abs}(a)$, and on the Planck mean opacity, $\kappa_{\rm P}(T,a)$ (see \citet{bjo01}, eq.~[5]),
\begin{displaymath}
T(a)^4 \propto {N_{\rm abs}(a) \over \kappa_{\rm P}(T,a)}.
\end{displaymath}
The number of absorbed photons is proportional to the absorption cross section averaged over the incident photons, so the equation above can be rewritten as
\begin{equation}
T(a)^4 \propto {\langle \kappa(a) \rangle \over \kappa_{\rm P}(T,a)} \ , \label{eq4}
\end{equation}
where 
\begin{equation}
\langle \kappa(a) \rangle \equiv \frac{\int \kappa_\lambda(a)J_\lambda d\lambda}{J} \ .
\end{equation}

The ratio $\kappa_\lambda(a)/\kappa_P $ is shown in Figure~\ref{sabs} for silicate grains with sizes ranging from $a=0.005$ to $1 \,\mu \rm m$, where the Planck mean opacity was calculated for $T = 1000\,\rm K$, the condensation temperature of silicate grains. This plot indicates the relative equilibrium temperature for different grain sizes as a function of the wavelength of the illuminating radiation field.
For example, if the grains are illuminated by a black body radiation of temperature $T_{\rm eff} = 2500\,\rm K$, which peaks at $\lambda = 1.2 \,\mu \rm m$, the grains with intermediate sizes ($a \approx 0.25$ to $0.50 \,\mu \rm m$) will be the hottest. Note, however, the range of temperatures will be relatively small in this case. 
On the other hand, if the grains are illuminated by a $20000 \rm K$ black body radiation field, the hottest grains will be the ones with $a \approx 0.05 \,\mu \rm m$ and all the other grain sizes will have much lower equilibrium temperatures, including the very small grains ($a = 0.005 \,\mu \rm m$). From eq.~(\ref{eq4}) and Figure~\ref{sabs}, the ratio between the maximum and minimum equilibrium temperature can be estimated to be $T_{\rm max} / T_{\rm min} \approx 2.5$. Hence, if the hottest grains have a temperature of $1000\,\rm K$, the coolest ones will be as cold as $400\,\rm K$.

This temperature difference between the different grain sizes is to be expected only for optically thin models. For optically thicker models, the temperature differences will be smaller, and for very optically thick envelopes the temperatures will be the same, regardless of the grain size. This occurs because at very large optical depths, $J_\lambda = B_\lambda$, and the grains are in thermal equilibrium with the radiation field.

We now can see an important difference between models of cool and hot stars. For cool stars, the equilibrium temperatures of the different grain sizes are not very different, so we expect the shape of the IR SED to be similar to those of the individual grain sizes. For hot stars, however, the spectral shape for each grain size will be very different. From the previous discussion, we infer that the IR SEDs of hot stars will depend on how the flux is split among the different grain sizes.

For an optically thin model, it is possible to estimate the values of $f(a)$, the wavelength integrated emission for grains of size $a$. From Kirchhoff's law, the emission must be the same as the wavelength integrated absorption, i.e.,
\begin{equation}
f(a) \propto \frac{dn}{da} \langle \kappa(a) \rangle \ . \label{fa}
\end{equation}
This relative emission is plotted in Figure~\ref{faf} for a standard MRN size distribution function, with $q = -3.5$. 
This figure shows the relative $f(a)$ as a function of the illuminating wavelength for the same grain sizes shown in Figure~\ref{sabs}. 
Note that, with the exception of the smallest grain considered, the curves of Figure~\ref{faf} scale approximately with the curves of Figure~\ref{sabs}, which means that, in general, the coolest grains will have the lowest integrated emission and vice-versa. As we will see later, this has an important consequence for the hot star case, for which the equilibrium temperature of the grains can be very different; however, because the hotter grains dominate the emission spectrum, the envelope IR SED will still be similar to the spectral shape of the individual grains.  

The smallest grains are an exception for this rule. If we compare them with the grains with $a \approx 0.05\,\mu\rm m$, we see that they reprocess about the same energy, but have lower equilibrium temperatures. This creates a situation in which grains with two different equilibrium temperatures both contribute equally to the final emission spectrum.

The above discussion can be summarized as follows. For cool stars, we expect the IR SED of the distribution to be similar to that of the single grain models because all the grains have similar temperatures and, hence, similar emission spectra. For hot stars, if we exclude the smallest grains $(\sim 0.005\,\mu \rm m)$, the IR SED of the distribution will be similar to the SED of the hottest grain, which dominates the emission. It is important to note, however, that these conclusions are valid only for the size distribution function assumed above. A different value of $q$, for example, will lead to changes in the relative values for $f(a)$.

\subsection{Approximate Invariance for Grain Size Distributions \label{sec52}}

In this section, we follow the same approach as section~\ref{single} and compare model SEDs for different grain size distributions.
More specifically, we employ a MRN size distribution function with $q=-3.5$, $a_{\rm min}=0.05 \,\mu \rm m$, and different values of $a_{\rm max}$. 
As before, we compare models with the same $\tau_{\rm rep}$, to ensure that condition 2 of AI is satisfied. 
As was seen in section~\ref{single}, different grain sizes can have different condensation radii. However, large grains typically form from smaller grains, so, as an approximation, we set the cavity radius to be the condensation radius of the smallest grain. In section~\ref{sec53}, we address the consequences and validity of this approach.
Finally, we defer discussion of the smallest grains ($a \approx 0.005 \,\mu \rm m$) until later, for the reasons mentioned in the last section. 

\subsubsection{Case A: Optically Thin Envelopes \label{casead}}

The case A results for models with a grain size distribution are essentially the same as case A for single grain size models.
If we compare model SEDs for models with $a_{\rm max} \lesssim 0.25 \,\mu \rm m$, the maximum grain size to satisfy condition 1 of AI, the curves are identical at the 3 to 5\% level, depending on the stellar temperature, as shown in Figure~\ref{sedad}.
This reinforces the conclusion for case A in section~\ref{caseas}, where we saw that the SED is a poor indicator of grain size (as long as the grain sizes present in the distribution each satisfy the condition of AI).

It is interesting to note that, this time, the smallest differences are observed for the hot star models, in contrast with the single grain size case, where the smallest differences were observed for cool star models. The reason for this can be understood from Figure~\ref{faf}, which shows that, for hot stars, a single grain size ($a \approx 0.05 \,\mu \rm m$) dominates the emission. In contrast, all grain sizes contribute to the cool star IR SED with slightly different emission spectra owing to small differences in the equilibrium temperatures. Another interesting point is that, if we include distributions with $a_{\rm max}$ larger than that satisfying condition 1 of AI, we still obtain very similar SEDs. For example, when $a_{\rm max} = 0.50 \,\mu \rm m$, the differences between the model SEDs is less than 10\%. This, again, contrasts with the results for single grain size models.

\subsubsection{Case B: Optically Thick Extinction with Optically Thin IR Emission  \label{casebd}}

Figure~\ref{sedbd} shows the results for models with $a_{\rm max} = 0.15$ and $0.25 \,\mu \rm m$ and optical depth set so that $\tau_{\rm rep}= 1$. Note the striking similarity in the extinction region, in contrast with the previous results for single grain size models (see Figure~\ref{sedbs}). The mean difference between the two SEDs is now only 11\% for $T_{\rm eff} = 2500\,\rm K$ and 10\% for $T_{\rm eff} = 20000\,\rm K$. The much closer similarity in the extinction region can be understood in terms of the extinction optical depth, shown in Figure~\ref{taubd}. As a result of the averaging of the optical properties with respect to the distribution function, the shape of the extinction optical depth of the two size distribution models is very similar.

The overall similarity of the SEDs for case B has important theoretical and observational consequences. From the theoretical point of view, this allows us to extend the consequences of AI from the IR SED to the entire spectrum, as was done for case A, but this time for a much broader range of optical depths.
From the observational point of view, this extends the conclusion that SED is a poor indicator of grain size, to include all optical depths.

\subsubsection{Case C: Optically Thick Envelopes also in the IR \label{casecd}}

This case, in all aspects, is similar to case C for single grain size models. Since the optical depth is very high, the grains tend to thermalize, approaching local thermal equilibrium, where no temperature difference is expected for the different grain sizes. Hence, the set of different grain sizes will behave as a single-sized grain whose optical properties are given by the average of the properties with respect to the distribution function.

\subsection{Breakdown of Approximate Invariance \label{sec53}}

The similarity of the SEDs for models with different size distributions for all optical depth regimes, illustrated in the last section, changes when we consider very small grains. 

Figure~\ref{sedbd2} shows the case B SEDs for models with $a_{\rm min} = 0.005\,\mu \rm m$ and $a_{\rm max} = 0.05, 0.15$ and $0.25 \,\mu \rm m$. As before, the models with large $a_{\rm max}$ have very similar SEDs in all spectral regions (compare Figure~\ref{sedbd2} with Figure~\ref{sedbd}); however, a significant difference in the extinction region arises for the model with $a_{\rm max} = 0.05\, \mu \rm m$. The reason for this difference is easily understood from the different shape of the extinction optical depth, shown in Figure~\ref{taubd2}.
This difference results from the fact that the smallest grains (in contrast with the larger ones) did not reach the geometrical limit throughout the extinction region.

\subsection{Breakdown of Grain Condensation Radius \label{sec54}}

In Figure~\ref{comp} we compare the SED for models with the same $a_{\rm max}$ ($0.25\,\mu \rm m$) and different  $a_{\rm min}$ (0.005 and $0.05\,\mu \rm m$) for a stellar temperature of $20000\,\rm K$. A large difference in the IR SED of the two models is observed. This difference arises from our choice of the cavity radius as the condensation radius of the smaller grains. For the model with $a = 0.05$ to $0.25\,\mu\rm m$ this choice is consistent, because the grains with $a \approx 0.05\,\mu\rm m$ are the hottest and they dominate the IR emission (see section~\ref{sec51}). For the model with $a = 0.005$ to $0.25\,\mu\rm m$, however, the smallest grain are cooler than the ones with $a \approx 0.05\,\mu\rm m$. From the choice of the inner radius, it follows that the equilibrium temperatures of the grains with $a \approx 0.05\,\mu\rm m$ is about $1250\,\rm K$, much higher than the grain condensation temperature, which is physically impossible. This causes the IR emission to shift to lower wavelengths, as seen in Figure~\ref{comp}.
It is evident that in this case a more consistent model should be considered, in which the intermediate (and larger) grains are not allowed to condense until their radiative equilibrium temperature drops below the condensation temperature.

Strictly speaking, this more consistent treatment should be used for all situations studied above, but it would demand a much more complicated procedure and would involve additional model parameters. The discussion of section~\ref{sec51} helps us set useful limits to when the approximation of identical condensation radii is valid for all grain sizes.
It will be approximately valid for cool stars, because the differences between the equilibrium temperatures of the grains are not large. It will be valid also for hot stars, when the smallest grain considered has the largest condensation temperature. Finally, the assumption fails for hot stars, when the smallest grain is cooler and has large integrated emission, comparable with hot grains, as seen above.

We can summarize this as follows: if all grain sizes meet the AI requirements, as is generally the case for cool stars, then the approximation of same condensation radius for all grain sizes is valid. If the grains do not meet the requirements, but a small range of grain sizes completely dominates the IR emission, then the approximate invariance of the SED is still maintained.
The more consistent treatment is required only when the grain sizes do not meet the AI requirements and different grain sizes have similar integrated emissivities. However, it is likely when one uses the correct condensation radius for each grain size that even in this last case AI will be recovered. 

\section{Summary \label{discussion}}

We have studied the effects of grain size on the SED of circumstellar envelopes with dust. To do so we introduced the concept of {\it approximate invariance}, as a very useful tool for revealing the essential similarities of the problem and for systematically exploring a large grid of model parameters.

The concept of AI follows from the fact that the optical properties of differently sized grains is similar in certain spectral regions, and from the idea that if a requirement for SED invariance is weakly violated (in this case, the shape of the grain opacity), the SED should still be approximately the same.
We studied separately single grain size models and models with a grain size distribution. For both situations, we studied three optical depths regimes: optically thin models ({\it case A}), optically thick models in the extinction region ({\it case B}) and very optically thick models ({\it case C}), for which most of the radiation emerges in the MIR to FIR. 

Our results for case A show that, given a grain composition and dust density distribution, the SED of models that comply with the conditions of AI are similar not only in the IR, but also in the extinction region. The spectral shape of the SED is controlled by a single parameter, the {\it reprocessing optical depth}, $\tau_{\rm rep}$, defined in eq.~(\ref{trep}). This parameter is {\it independent of grain size}. Our results for case C are similar but, as shown in section~\ref{casecs}, another parameter ($\Gamma$, defined in eq.~(\ref{gamma})) controls the shape of the SED. 

The results for case B differ for single grain size models and models with a size distribution. For single grain size models, the SEDs are similar only in the IR, while large differences are observed in the extinction region, where the grain opacities depend very much on grain size. Models with a grain size distribution, on the other hand, display a striking similarity in the extinction region, provided the lower limit of the grain size distribution, $a_{\rm min}$, is in the geometrical limit in the extinction region. It follows that, if $a_{\rm min}$ is chosen accordingly, the case B SEDs for models with a grain size distribution is controlled solely by the parameter $\tau_{\rm rep}$ (the same result as case A).

The fact that $\tau_{rep}$ (or $\Gamma$ for case C) is the appropriate parameter for SED classification indicates that the usual approach in the literature - using $\tau_V$ or $\tau_F$ - will not reveal the general invariance of the SED discussed in this paper. We conclude therefore that the SED is generally a very poor constraint of grain size. In most situations, observations must resolve differences of the order of a few percent to extract information about the grain size. If we add to this the fact that other model parameters, such as dust spatial distribution, composition and optical properties are somewhat uncertain, it becomes apparent that the task of extracting information about grain sizes from the SED alone may prove very difficult or even impossible for spherical geometries.

If follows from AI that in most circumstances $\tau_{\rm rep}$ is the {\it only parameter related to the grain opacity  that can be unambiguously extracted from the observations}. 
Using $\tau_{\rm rep}$ to determine dust column density or mass is directly subject to the uncertainty in the determination of the grain sizes, which directly affects our ability to measure mass loss rates using the SED.

\begin{figure}[bp]
\plotone{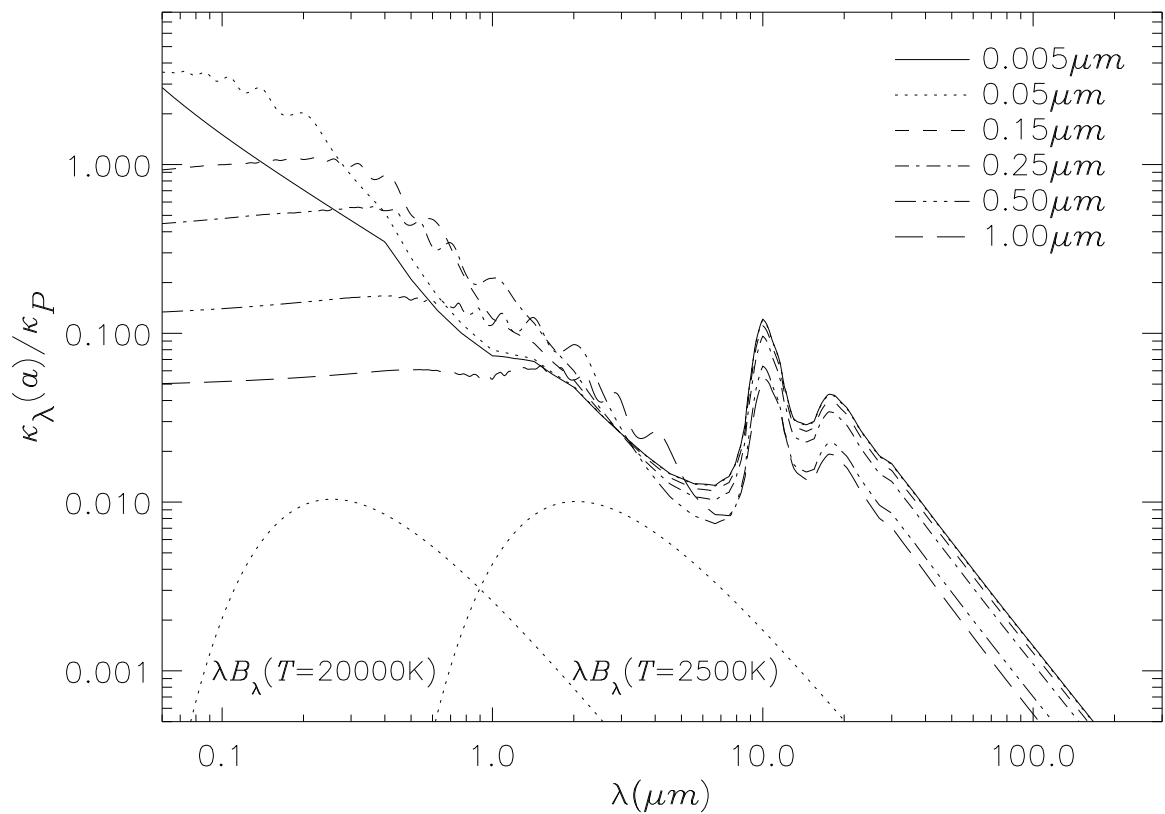}
\caption{$\kappa_{\lambda}(a)/\kappa_{\rm P}$ vs. wavelength.
Shown are curves for silicate grains with different sizes, as indicated. Also shown is the black body spectrum for $T_{\rm eff} = 2500$ and $20000\,\rm K$. The Planck mean opacity was calculated for $1000\,\rm K$. The unit of $\lambda B_\lambda$ is arbitrary. \label{sabs}}
\end{figure}

\begin{figure}[bp]
\plotone{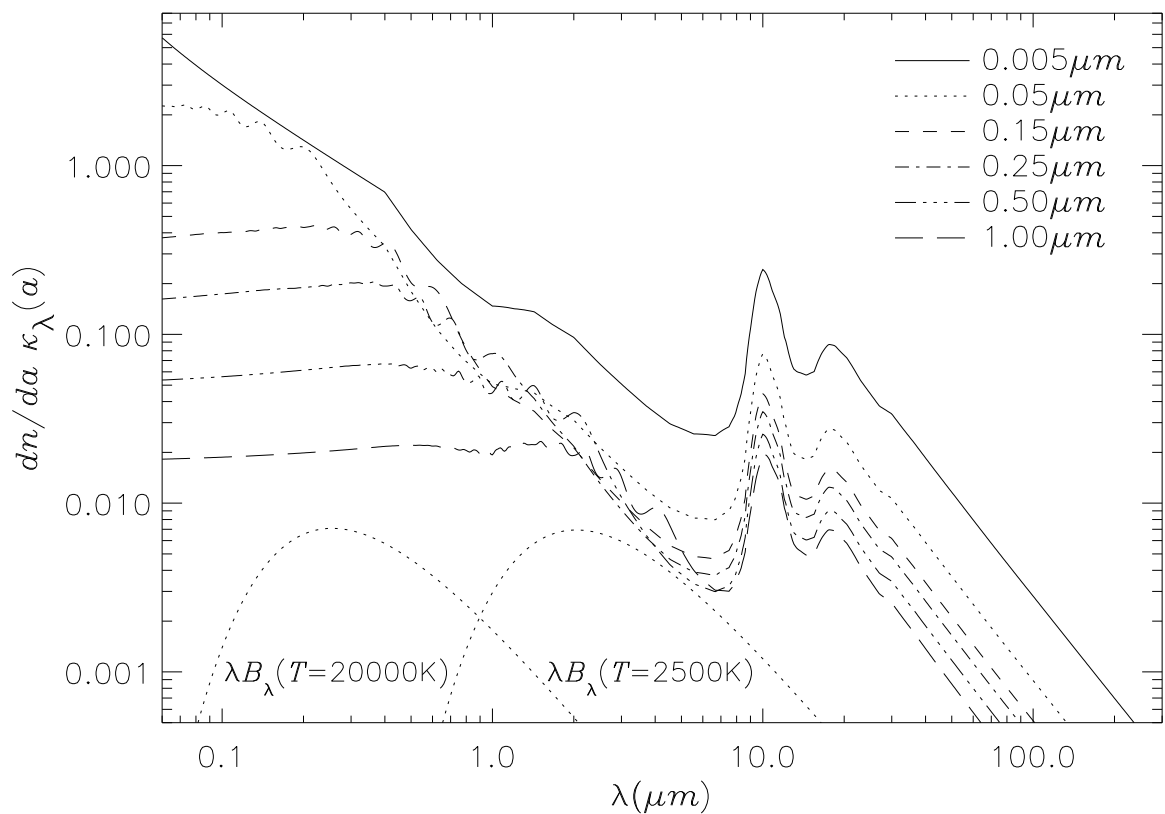}
\caption{$(dn/da)\,\kappa_{\lambda}(a)$ vs. wavelength.
Shown are the curves for silicate grains with different sizes, as indicated. Also shown is the black body spectrum for $T_{\rm eff} = 2500$ and $20000\,\rm K$. The size distribution $dn/da$ is a power-law, with index $q=-3.5$. The units of $(dn/da)\,\kappa_{\lambda}(a)$ and $\lambda B_\lambda$ are arbitrary. \label{faf}}
\end{figure}

\begin{figure}[bp]
\plottwo{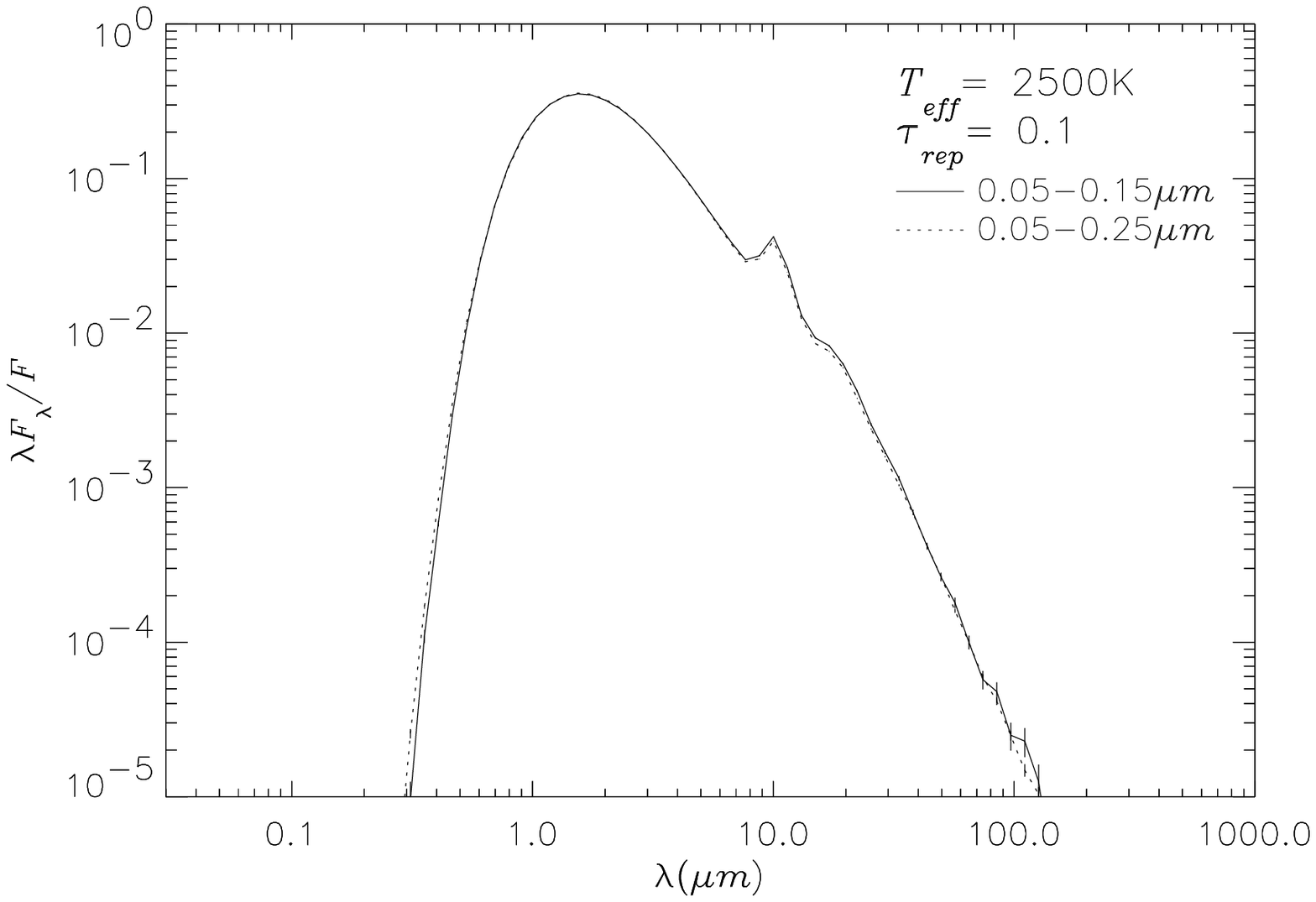}{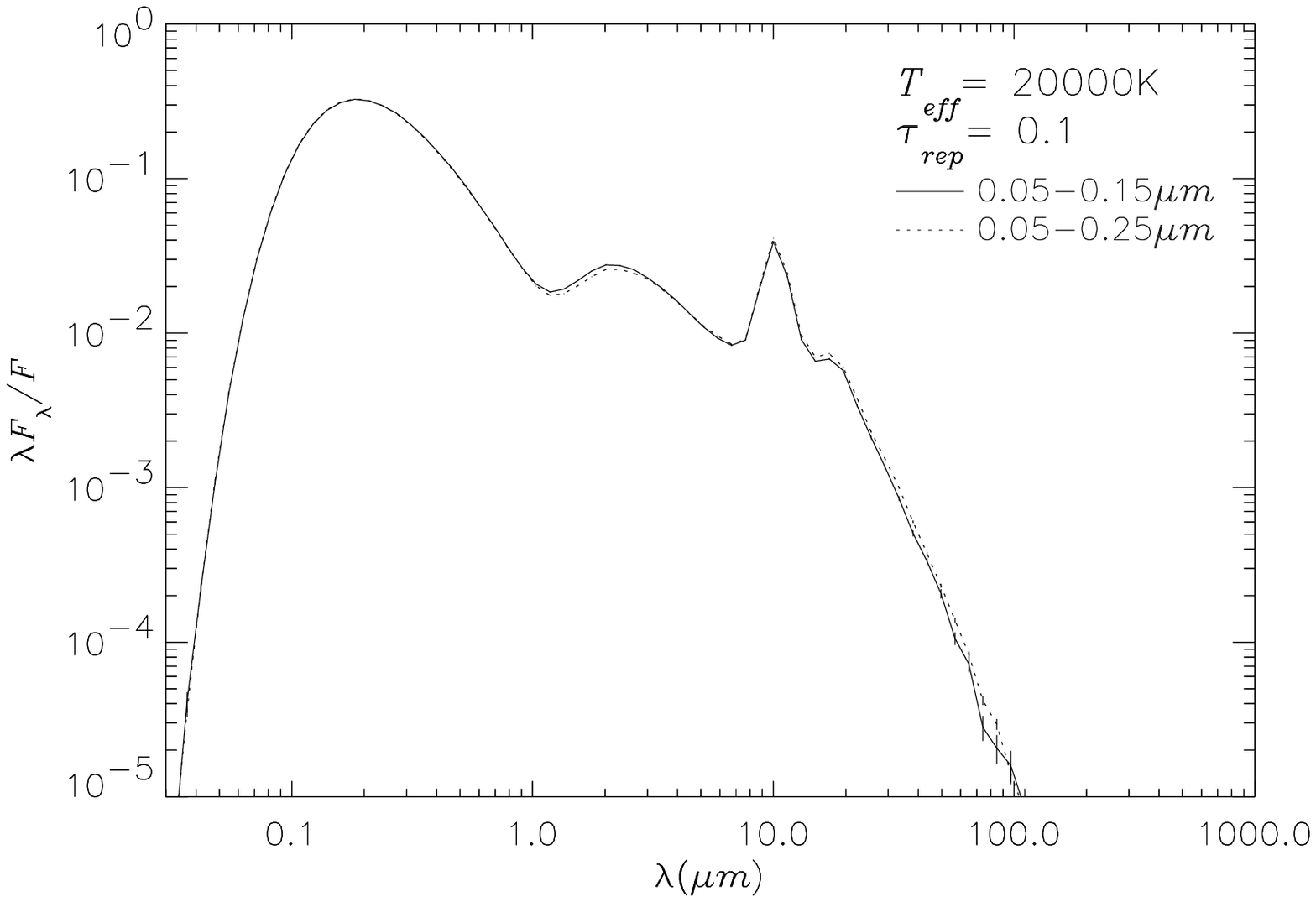}
\caption{Size distribution models for case A, with $T_{\rm i} = 1000\,\rm K$. Shown are the results for two envelope models of cosmic silicate grains with a MRN size distribution and size range as indicated. The left and right panels show the results for a given stellar temperature. For each model, the optical depth was adjusted so that $\tau_{\rm rep}=0.1$. \label{sedad}}
\end{figure}

\begin{figure}[bp]
\plottwo{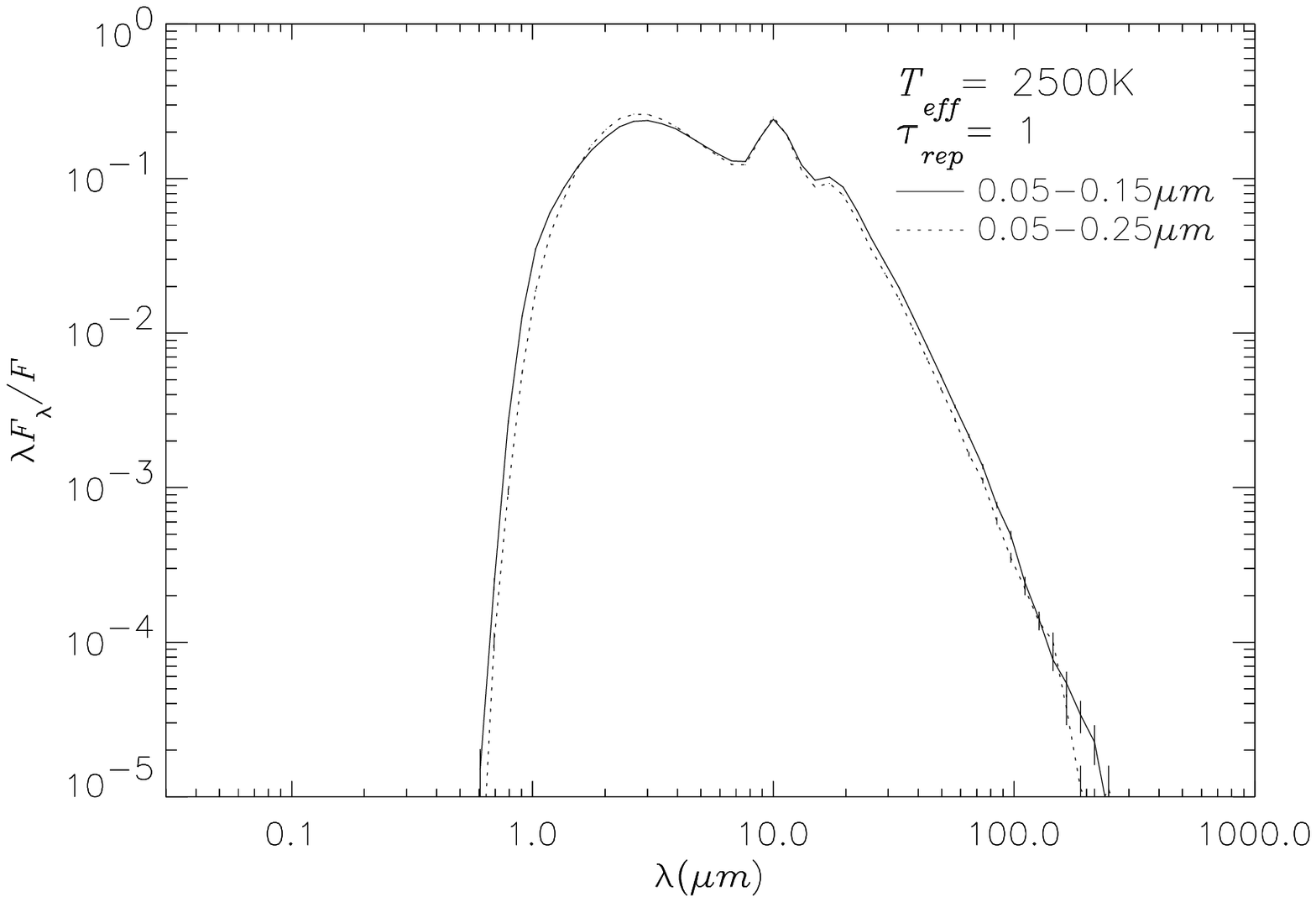}{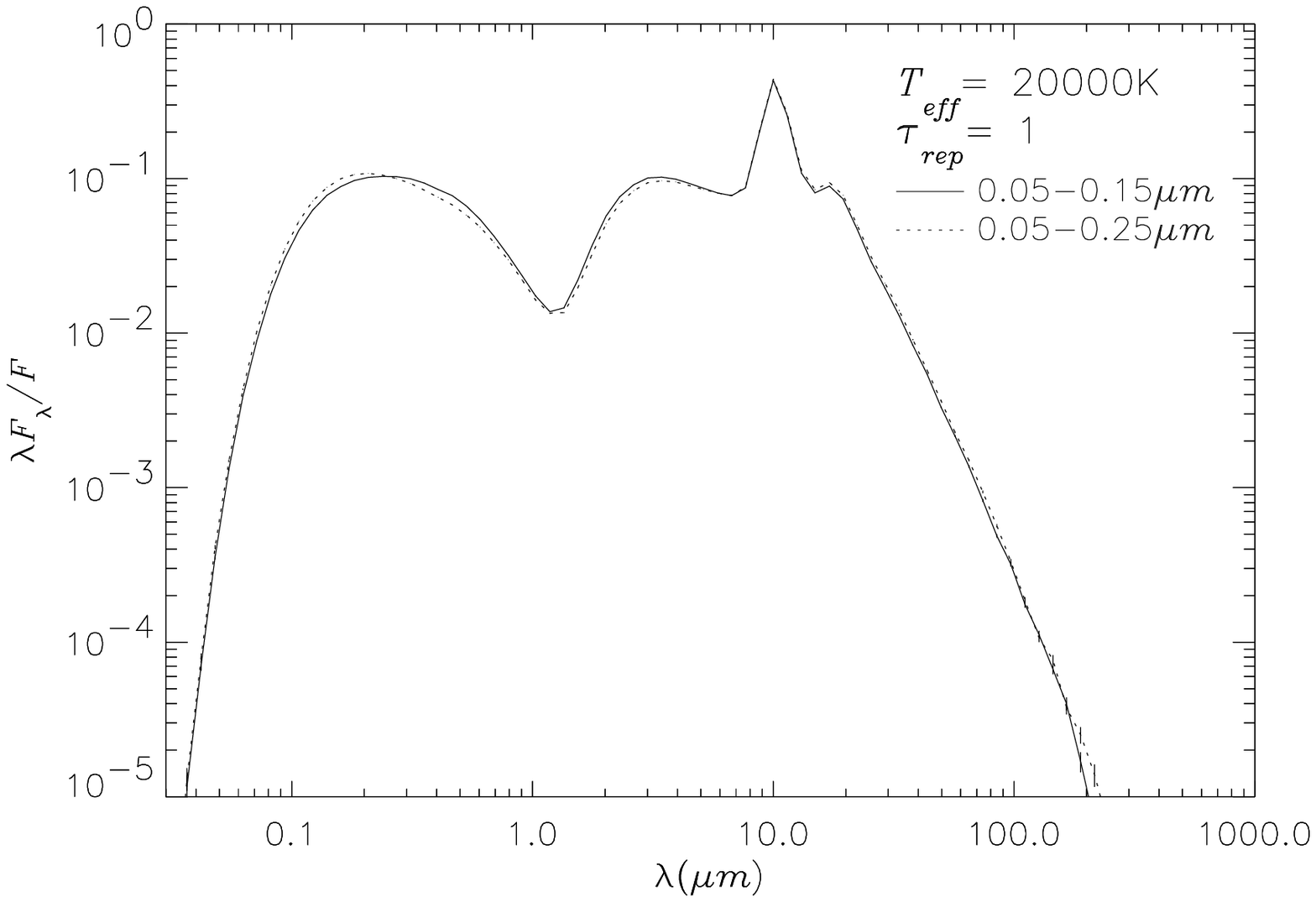}
\caption{Size distribution models for case B, with $T_{\rm i} = 1000\,\rm K$. Shown are the results for two envelope models of cosmic silicate grains with a MRN size distribution and size range as indicated. The left and right panels show the results for a given stellar temperature. For each model, the optical depth was adjusted so that $\tau_{\rm rep}=1$. \label{sedbd}}
\end{figure}

\begin{figure}[bp]
\plotone{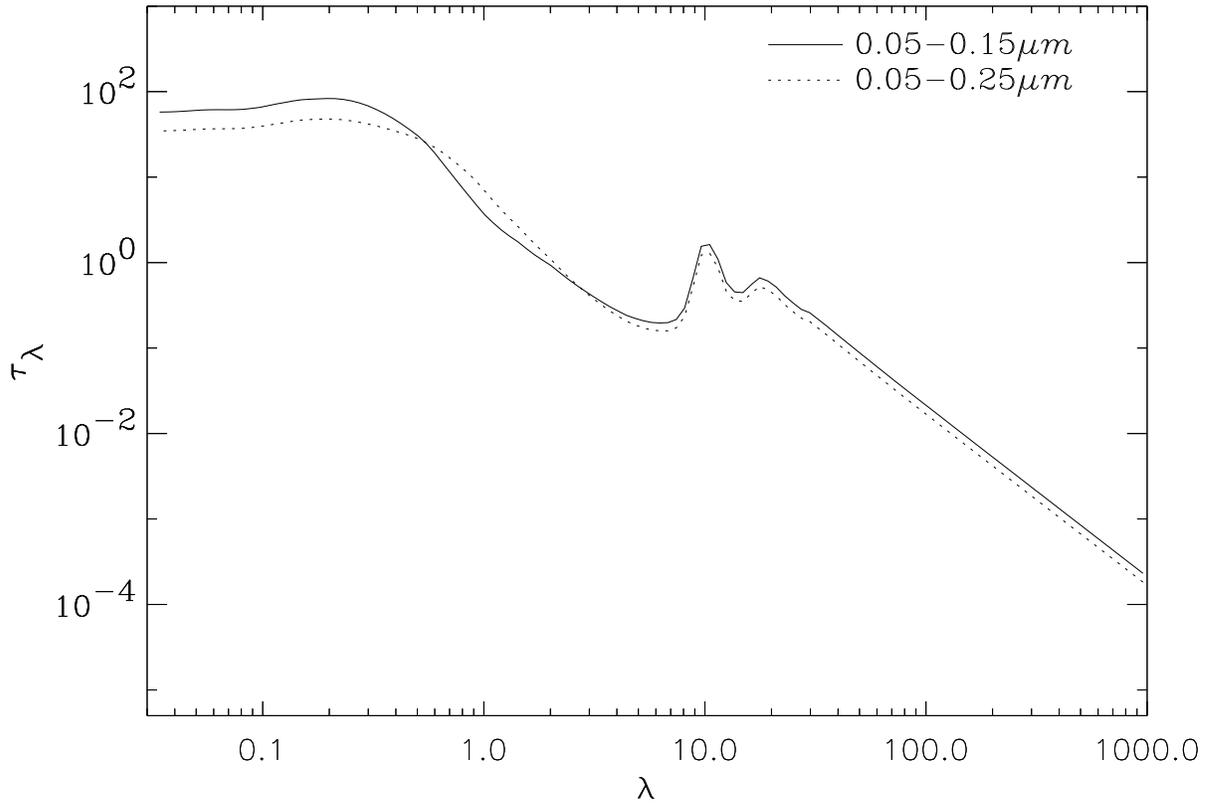}
\caption{Extinction optical depth, $\tau_\lambda$, vs. wavelength for the $2500 \rm K$ models shown in Figure~\ref{sedbd}. Note the similar shape in the optical. \label{taubd}}
\end{figure}

\begin{figure}[bp]
\plottwo{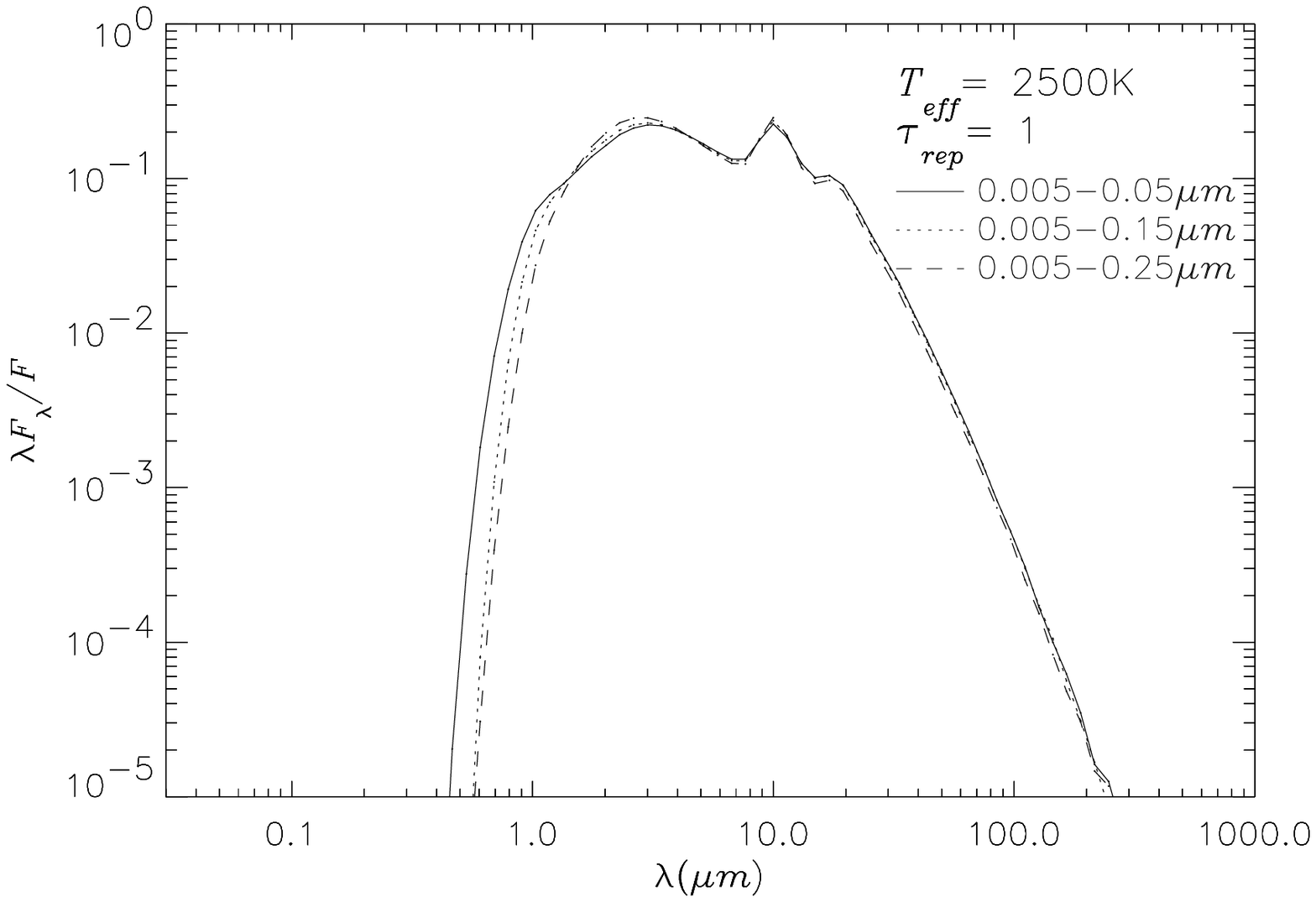}{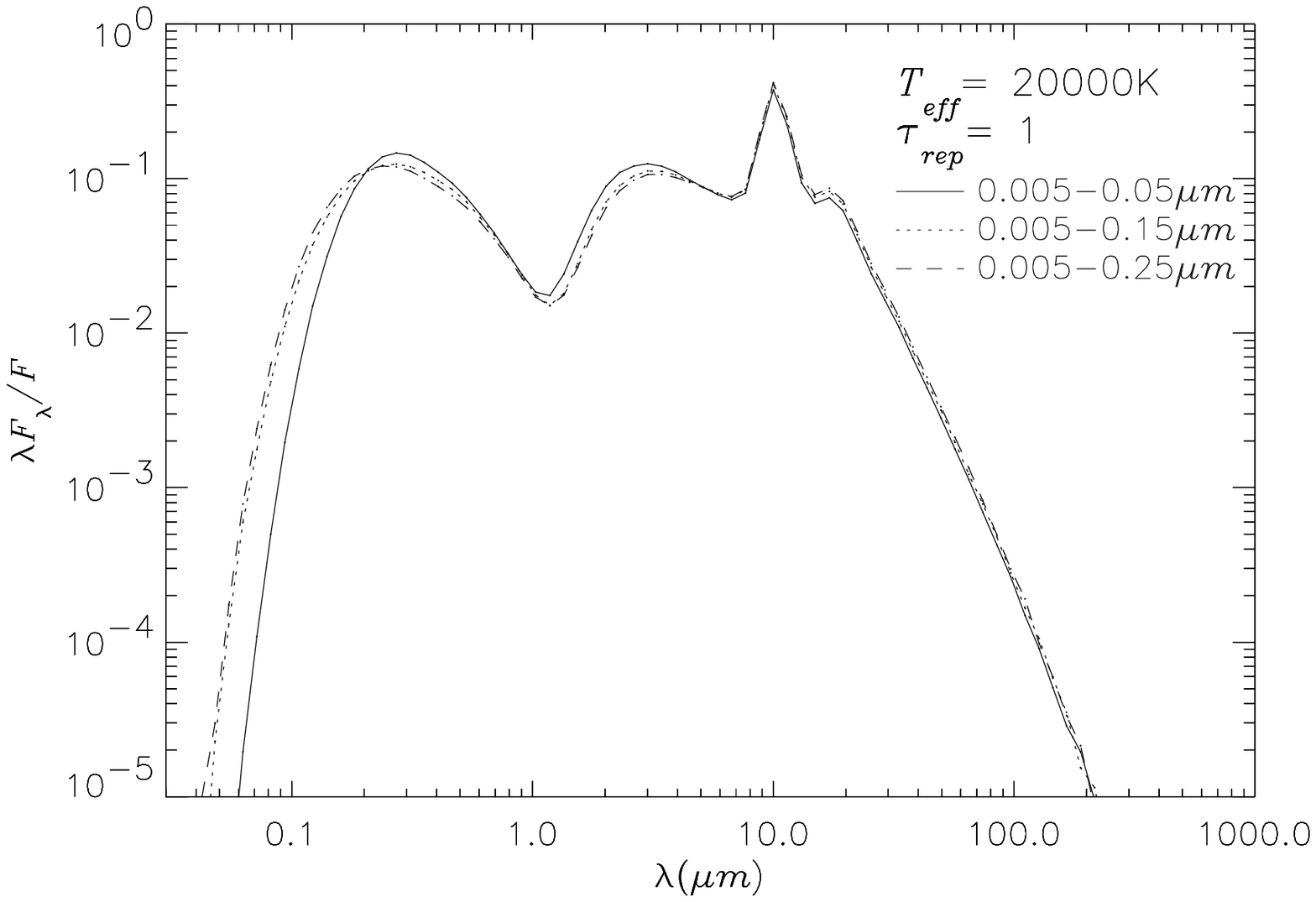}
\caption{Size distribution models for case B, with $T_{\rm i} = 1000\,\rm K$. Shown are three envelope models of cosmic silicate grains with a MRN size distribution and size range as indicated. The left and right panels show the results for a given stellar temperature. For each model, the optical depth was adjusted so that $\tau_{\rm rep}=1$. \label{sedbd2}}
\end{figure}

\begin{figure}[bp]
\plotone{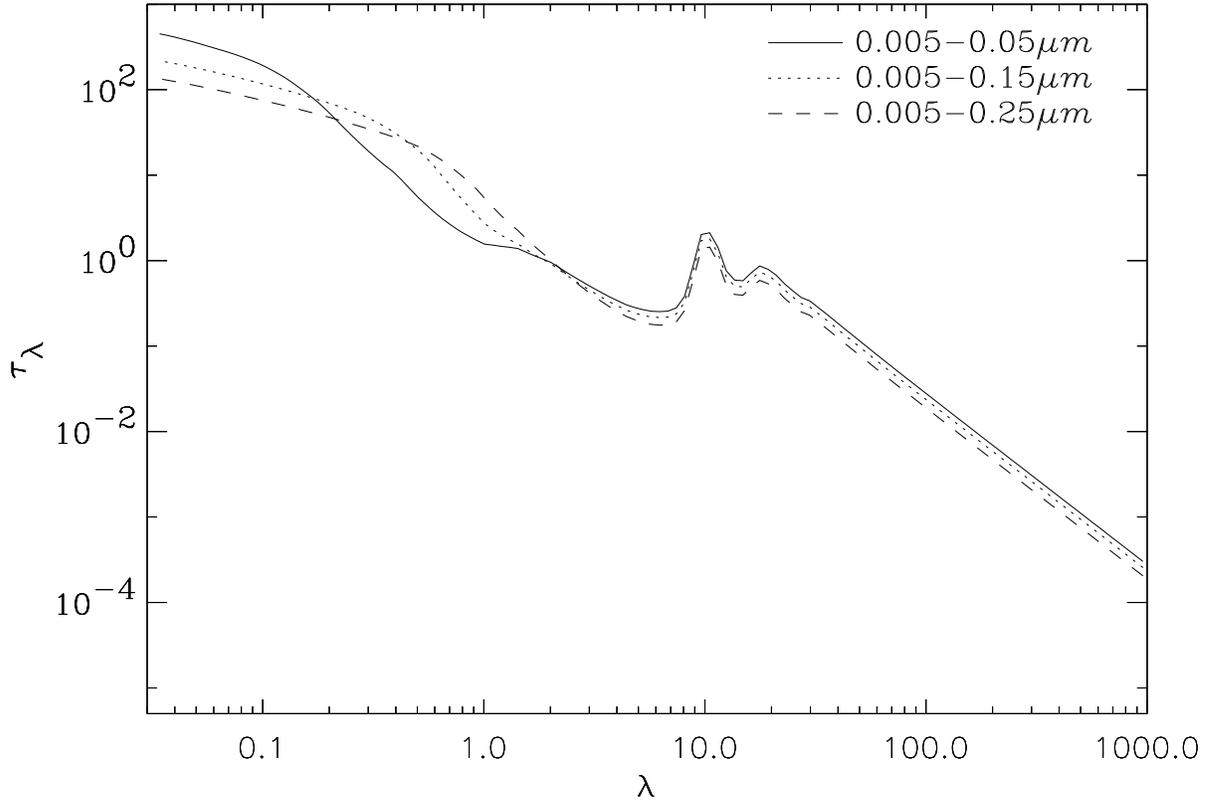}
\caption{Extinction optical depth, $\tau_\lambda$, vs. wavelength for the $2500 \rm K$ models shown in Figure~\ref{sedbd2}. \label{taubd2}}
\end{figure}

\begin{figure}[bp]
\plotone{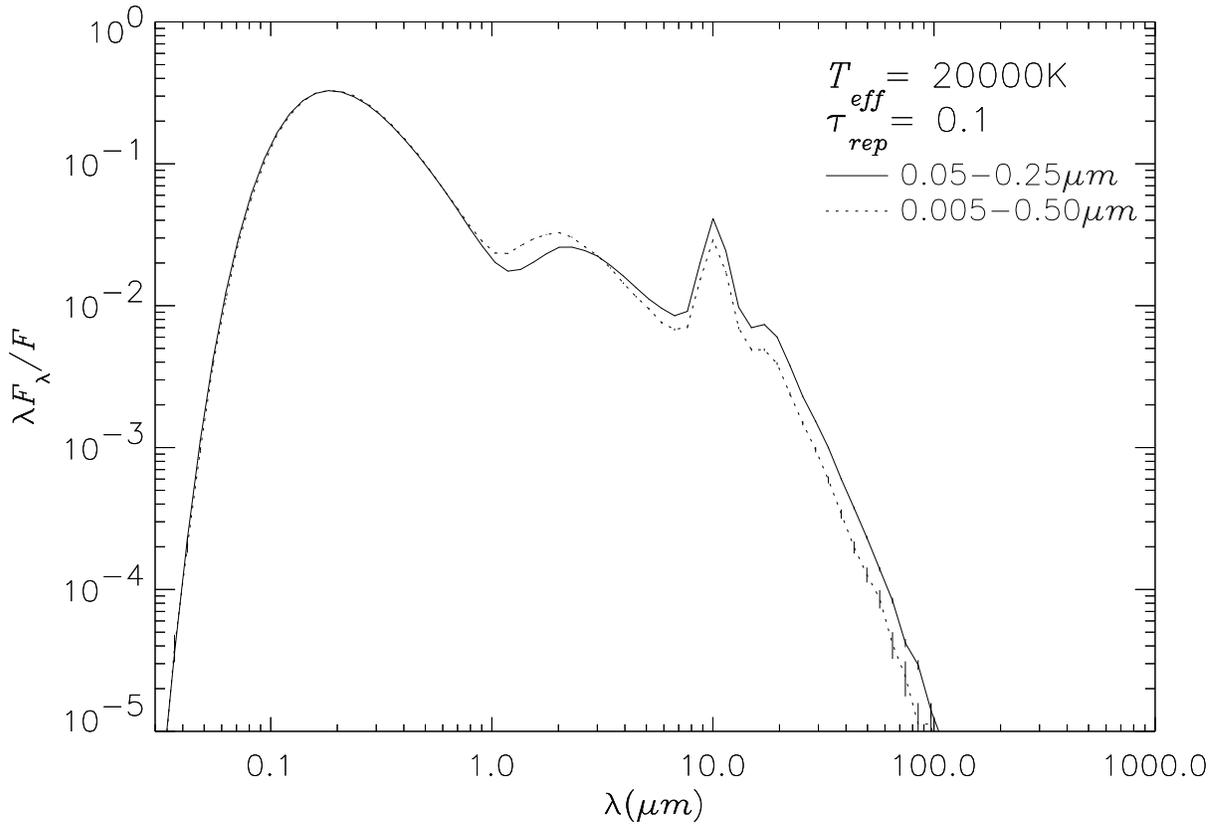}
\caption{SEDs for size distribution models. The figure compares the SED for a model with $a = 0.005$ to $0.25\,\mu\rm m$ with the SED for a model with $a = 0.05$ to $0.25\,\mu\rm m$. Both models have $\tau_{\rm rep} = 0.1$ and $T_{\rm i} = 1000\,\rm K$. \label{comp}}
\end{figure}

An important physical effect discussed in section~\ref{sec51} is that for models with a grain size distribution, the different grain sizes have different radiative equilibrium temperatures. This implies that a consistent model should include the correct condensation radius for each grain size of the distribution. However, we show in section~\ref{sec51} that the approximation of using identical condensation radii is reasonable when either all grain sizes have similar condensation temperatures or when a particular range of grain sizes dominate the IR emission. The only situation where the consistent treatment (each grain size having its correct condensation radius) is required is when two or more ranges of grain sizes have different equilibrium temperatures and similar integrated emission. This situation was found only for models of hot stars with very small grains ($a \lesssim 0.01\,\mu \rm m$) present in the distribution. 
It is important to notice that the results presented here depend on the particular choice of the distribution function, because this will control the relative contribution of each grain size to the emission.
We suggest that a study of a given system could start with an analysis of the chosen dust size distribution along the lines described in section~\ref{sec51}, Figures~\ref{sabs} and \ref{faf}. Such analysis can be very useful because one would know beforehand if the approximation of identical condensation radii for all grain sizes of the distribution is a valid one.

We have presented, here, results only for silicate grains. However, this study was carried out for amorphous carbon and silicon carbide grains as well. The results for these materials are not shown because, although the specific details of the results are different (see \citet{car01}), the fundamental concepts discussed here remain valid.

\acknowledgments{A.C.C. and A.M.M. acknowledge support from S\~ao Paulo State Funding Agency (grants 98/02238-2 and 01/12589-1). A.C.C. and J.E.B. acknowledge support from NSF grant AST-9819928. A.M.M. acknowledges partial support from CNPq.}
\acknowledgments{The authors wish to thank the referee for his/her very useful comments.}

\clearpage

\clearpage

\begin{deluxetable}{cccc}
\tablecaption{Inner cavity radius, grain masses and  $V$-band optical depth for models of Figure~\ref{sedas}. \label{tab1}}
\tablewidth{0pt}
\tablehead{
\colhead{$a\,(\mu \rm m)$} &
\colhead{$r_{\rm i} (R_\star)$} &
\colhead{$M/M_{a=0.005\,\mu \rm m}$\tablenotemark{*}} &
\colhead{$\tau_{V}$}
}

\startdata
\cutinhead{$2500\,\rm K$}
0.005 & 4.5  & 1.00 & 0.308\\
0.05  & 4.64 & 0.97 & 0.640\\
0.15  & 5.62 & 0.85 & 2.98  \\
0.25  & 6.26 & 0.74 & 1.54  \\
\cutinhead{$20000\,\rm K$}
0.005 & $1.00 \cdot 10^3$ & 1.00 & $2.32 \cdot 10^{-2}$\\
0.05  & $1.53 \cdot 10^3$ & 1.00  & $2.25 \cdot 10^{-2}$ \\
0.15  & $1.12 \cdot 10^3$ & 0.92  & 0.300                           \\
0.25  & $8.12 \cdot 10^2$ & 0.78  & 0.358
\enddata
\tablenotetext{*}{$M/M_{a=0.005\,\mu \rm m}$ is the dust mass normalized to the dust mass of the model with $a=0.005\,\mu \rm m$}
\end{deluxetable}

\clearpage

\clearpage

\begin{deluxetable}{cccccc}
\tablecaption{Inner cavity radius, grain masses and  $V$-band optical depth for models of Figure~\ref{sedbs}. \label{tab2}}
\tablewidth{0pt}
\tablehead{
\colhead{$a\, (\mu \rm m)$} &
\colhead{$r_{\rm i} (R_\star)$} &
\colhead{$M/M_{a=0.005\,\mu \rm m}$} &
\colhead{$\tau_{V}$}
}

\startdata
\cutinhead{$2500\,\rm K$}
0.005 & 4.9 & 1.00 & 3.49 \\
0.05  & 5.3 & 1.09 & 7.42  \\
0.15  & 7.7 & 1.48 & 37.0  \\
0.25  & 9.2 & 1.30 & 17.0  \\
\cutinhead{$20000\,\rm K$}
0.005 & $1.04 \cdot 10^3$ & 1.00 & 0.273\\
0.05  & $1.70 \cdot 10^3$  & 0.96  & 0.224\\
0.15  & $1.17 \cdot 10^3$  & 0.74  & 2.81  \\
0.25  & $8.61 \cdot 10^2$  & 0.65  & 3.41  
\enddata

\end{deluxetable}

\clearpage

\clearpage

\begin{deluxetable}{cccc}
\tablecaption{Inner cavity radius, grain masses and optical depth at $100 \,\mu \rm m$ for the models shown in Figure~\ref{sedcs2}. \label{tab3}}
\tablewidth{0pt}
\tablehead{
\colhead{$a\, (\mu \rm m)$} &
\colhead{$r_{\rm i} (R_\star)$} &
\colhead{$M/M_{a=0.005\,\mu \rm m}$} &
\colhead{$\tau_{100}$}
}

\startdata
0.005 & $1.75 \cdot 10^3$ & 1.00 & 5.2\\
0.05 & $1.77 \cdot 10^3$ & 0.99 & 5.1\\
0.15  & $1.85 \cdot 10^3$ & 1.03 & 4.8\\
0.25  & $1.97 \cdot 10^3$ & 1.04 & 4.3\\
0.50  & $2.4 \cdot 10^3$ & 1.37 & 3.8
\enddata

\end{deluxetable}

\clearpage

\begin{deluxetable}{cccccc}
\tablecaption{ $V$-band optical depth for anisotropic SPF ($\tau_{V}$) and isotropic SPF ($\tau_{V}^{\rm iso}$), for case A ($\tau_{\rm rep}=0.1$) and case B ($\tau_{\rm rep}=1$) models. Also shown is the flux mean optical depth ($\tau_{\rm F}$) for the models with anisotropic SPF. \label{tab4}}
\tablewidth{0pt}
\tablehead{
\colhead{$\tau_{V}$ (case A)} &
\colhead{$\tau_{V}^{\rm iso}$ (case A)} &
\colhead{$\tau_{\rm F}$ (case A)} &
\colhead{$\tau_{V}$ (case B)} &
\colhead{$\tau_{V}^{\rm iso}$ (case B)} &
\colhead{$\tau_{\rm F}$ (case B)}
}

\startdata
\cutinhead{$2500\,\rm K$}
0.308 & 0.307  & $1.01\cdot 10^{-1}$ & 3.49   & 3.47  & 1.15\\
0.640 & 0.647 & $9.96\cdot 10^{-2}$  & 7.42    & 7.44    & 1.16\\
2.98   & 2.83 & $9.48\cdot 10^{-2}$     & 37.0    & 35.7  & 1.18\\
1.54   & 1.39 & $9.45\cdot 10^{-2}$     & 17.0    & 15.7  & 1.04\\
\cutinhead{$20000\,\rm K$}
$2.32 \cdot 10^{-2}$  & $2.32 \cdot 10^{-2}$ & $1.02\cdot 10^{-1}$ & 0.273  & 0.272  & 1.20\\
$2.25 \cdot 10^{-2}$   & $2.09 \cdot 10^{-2}$ & $9.83\cdot 10^{-2}$& 0.224 & 0.183  & $9.79\cdot 10^{-2}$\\
0.300                             & 0.284                           & $1.01\cdot 10^{-1}$& 2.81   & 2.31     & $9.46\cdot 10^{-2}$\\
0.358                             & 0.346                           & $9.96\cdot 10^{-2}$& 3.41   & 2.83     & $9.49\cdot 10^{-2}$
\enddata
\end{deluxetable}

\clearpage

\clearpage

\clearpage

\end{document}